\documentclass[aps,prc,preprint,showpacs]{revtex4}
\usepackage[T1]{fontenc}
\usepackage[latin1]{inputenc}
\usepackage{graphics}

\begin{document}

\newcommand{\bd}[1]{ \mbox{\boldmath $#1$}  }
\newcommand{\xslash}[1]{\overlay{#1}{/}}
\newcommand{\sla}[1]{\xslash{#1}}
\thispagestyle{empty}  
  
\begin{center}  
  
{\Large \bf One dimensional scattering of a two body interacting system 
by an infinite wall}  

\vspace{0.5cm} 
{\large A.M. Moro, J.A. Caballero and J. G\'omez-Camacho} 
 
\date{\today}
 
\vspace{.3cm} 
{Departamento de FAMN. Facultad de F\'{\i}sica. Universidad de Sevilla.\\
Apdo. 1065, 41080 Sevilla (Spain)}\\
 
\end{center} 
 
\vspace{0.5cm} 

\begin{abstract}
The one-dimensional scattering of a two body  interacting system  
by an infinite wall is studied in a quantum-mechanical framework. 
This problem contains some
of the dynamical features present in the collision of atomic, molecular and 
nuclear systems. The scattering problem is solved exactly, for the case of a 
harmonic interaction between the fragments. The exact result is used to assess
the validity of two different approximations to the scattering process.
The adiabatic approximation, which considers that the relative co-ordinate 
is frozen during the scattering process, is found to be inadequate for this
problem. The uncorrelated scattering approximation, which neglects the 
correlation between the fragments, gives results in accordance with the
exact calculations when the scattering energy is high compared to the
oscillator parameter.
\end{abstract}
 


\maketitle
\section{Introduction\label{section:intro}}

Recently, important efforts in the fields of molecular, atomic and nuclear
physics have been devoted to the analysis of collision processes involving 
composite quantum systems. Despite the peculiarities of the different fields,
the theoretical description of the collision has several common features.
R-matrix theory \cite{Fano86} is used for re-arrangement collisions, when
several electrons or nucleons may be exchanged between the colliding systems.
Coupled channels calculations \cite{Fano86,Murr89} are used to describe 
excitation and dissociation in atomic, molecular and nuclear systems. The
``adiabatic'' or ``sudden'' approximation is often invoked in molecular and 
nuclear collisions, because it simplifies significantly the description of
the scattering process, by considering that some of the relevant co-ordinates 
are effectively frozen during the scattering process. Also, the description of
 the collision of atomic and molecular beams with surfaces requires 
coupled channels descriptions, but the combined difficulties of the 
atom-surface interactions and molecular vibrations makes the ``sudden'' 
approximation almost essential for its solution \cite{Murr89}.

In the case of atomic and molecular
physics, one-dimensional atom-molecule collision reactions have been
studied in detail by solving the Schr\"odinger equation~\cite{Levi87}. In
particular, collisions between an atom and a diatomic molecule represented
by harmonic~\cite{Rapp60}, an-harmonic~\cite{Reca87} and Morse 
oscillators~\cite{Wulf83} have been analysed with different degrees of 
approximation. Diverse computational methods have been also implemented
to study three-dimensional molecular and atomic collisions~\cite{Zhan88}.
Moreover, in some recent papers algebraic approaches have been proposed
for describing one- and three-dimensional atom-molecule collision 
processes~\cite{Sant94,Fran92}.

In the field of nuclear physics, much of the interest
in recent years has been focused on 
the study of the properties of halo nuclei, 
weakly bound systems characterised by the existence
of one or two particles (generally neutrons) with a high probability of being
at distances larger than the typical nuclear radius. 
Different approaches have been used in the analysis of reactions involving
halo nuclei. The adiabatic approximation~\cite{Ron70,Ron97,Ron97b,Tos98}
assumes that, for sufficiently high scattering energies, the
internal Hamiltonian is accurately represented by its corresponding eigenvalue
for the ground state.
The sudden approximation~\cite{Edu96,Edu97,Edu98a,Edu98b}
relies on two main assumptions: i) the impulse approximation, i.e., 
the multiple scattering
expansion for the T-matrix is approximated by the sum on the 
individual T-matrices
for the scattering of the separated constituents, and ii) only one 
of the particles of the projectile interacts with the target. 
In the high energy regime, the Glauber approach~\cite{Gla59}, which combine
eikonal dynamics with the adiabatic approximation, 
provides a simple tool to analyse reactions involving halo nuclei. 
We have recently developed an alternative approach to 
the description of weakly bound systems. The approach, called ``Uncorrelated
Scattering Approximation'' (USA)~\cite{Mor00,Mor01},
is based on the fact that, for a weakly bound
projectile, the correlations between the constituents are weak and so, to some
extent, they are expected to evolve independently in the 
strong field of a heavy target. Thus, the three body S-matrix can be 
expressed in terms of the individual
two body S-matrices for the scattering of the constituents. In this framework,
the scattering observables of the process are given in terms of 
two body constituent-target observables. 
This model has been applied to describe
elastic scattering and break-up of deuterons on heavy targets, with encouraging
results. The Uncorrelated Scattering Approximation has certain relation with
the R-matrix approach. In a R-matrix calculation of deuteron scattering, 
the wave-function within the range of the
target interaction is given in terms of products of
single-particle (protons and neutrons) wave-functions 
with the proper boundary conditions, provided the interaction between the
fragments of the projectile is neglected. These single-particle
wave-functions should be matched with the proper asymptotic 
wave-functions. In the USA calculation, the incident wave is expanded in terms
of products of fragment-target wave-functions, which then scatter independently
from the target. 

The objective of this work is to investigate the
scattering of a composite system from a target with
interactions which have a very short range compared not only with
the size of the
system, but also with the associated wave-length of the projectiles. For that
purpose, we consider the case of two particles, interacting through a
harmonic oscillator potential, which collide with an infinite wall. We develop
two different methods to solve the problem exactly, obtaining the wave-function
as well as the S- matrix, or reflection coefficients, which give the 
probability amplitudes for the excitation of the different oscillator states.
We compare the exact results with the ones obtained using the adiabatic 
approximation, and the uncorrelated scattering approximation.

The model discussed here does not pretend to be a realistic representation
of any specific molecular, atomic or nuclear system. However, it
has the virtue that the only length scale is the
oscillator length $a_0$, while the only energy scale is the oscillator
parameter $\hbar \omega$. Then, the results obtained, which are
expressed in terms of $r/a_0$ and $E/\hbar \omega$, can be applied 
in principle
to arbitrary energy  and length scales, which may be nuclear, atomic or
molecular. This fact makes the model attractive as a bench-mark to
test the validity of the approximations which are used for the description of
composite systems.

One aspect of the model which seems odd is the infinite nature of the 
harmonic oscillator interaction between the fragments. This would prevent
dissociation in the case of molecules, ionization in the case of atoms or
break-up for nuclei. On the other hand, the harmonic oscillator basis is 
complete, and thus the results of the model incorporate
effects on the scattering due to coupling to all possible, open or closed, 
states.

The paper is organised as follows: in section~II we describe
the problem to be solved. Two different approaches to extract the 
exact solutions are developed. In section~III 
we derive the S-matrix in the adiabatic
model approach. In section~IV we present the Uncorrelated
Scattering Approximation. In section~V we apply it to evaluate 
the scattering coefficients
and compare with the exact solution. Section~VI is devoted
to summary and conclusions.

\begin{figure}
{\par\centering \resizebox*{0.45\textwidth}{!}{\includegraphics{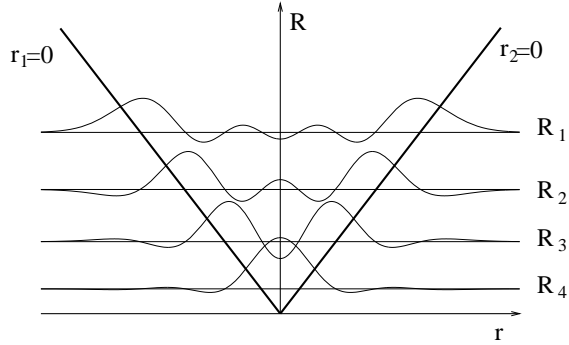}}\par}
\caption{\label{Fig:exact_cls}Schematic plot representing the scattering problem in
the \protect\( (r,R)\protect \) plane. The waves represent the SCLS associated
with a basis with \protect\( N=8\protect \) HO states. Each one of these SCLS
has been plotted at the center of mass distance at which it is supposed to be
scattered by the wall, according to the formalism described in section \ref{sec:exact_CLS}.}
\end{figure}

\begin{figure*}
{\par\centering \resizebox*{0.65\textwidth}{!}{\includegraphics{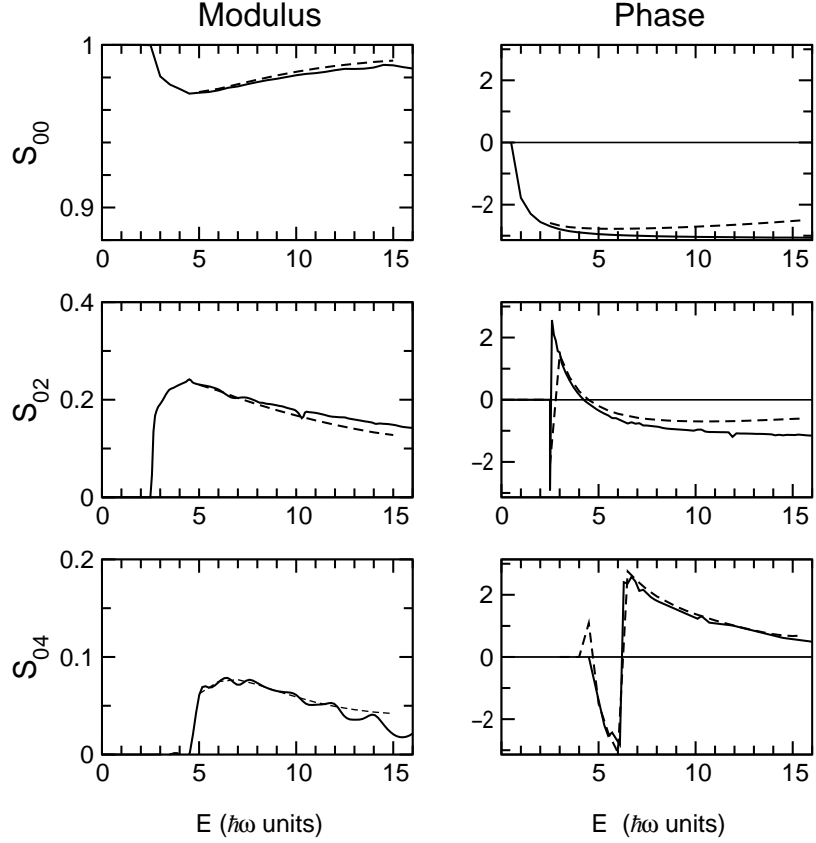}} \par}
\caption{\label{Fig:Sexact(E)}S-matrix elements versus the scattering 
energy for the
HO states \protect\( n=\protect \)0, 2 and 4. The solid and dashed lines 
correspond
to the exact calculation using the methods of sections~\ref{sec:exact_direct}
and~\ref{sec:exact_CLS}, respectively. A basis space with 15 even parity HO
states was used for the calculations.}
\end{figure*}

\section{Statement of the problem and exact solution\label{sec:problem}}

Let us consider the scattering of a two-particle bound system by 
the short range
potential due to a heavy target, placed at the origin of coordinates.

The Hamiltonian for the system may be written in terms of the coordinates of
the two particles (\( r_{1} \), \( r_{2} \)) and their momenta (\( p_{1} \),
\( p_{2} \)) as

\begin{eqnarray}
\hat{H} & = & \hat{h}_{1}+\hat{h}_{2}+v(r_{1}-r_{2})\label{eq:Hr1r2} \\
\hat{h}_{i} & = & \frac{\hat{p}_{i}^{2}}{2m_{i}}+v_{i}(r_{i})\label{eq:hi} 
\end{eqnarray}
where \( m_{i} \) are the masses of the constituents, \( v_{i}(r_{i}) \)
is the potential exerted by the target on each particle 
and \( v(r_{1}-r_{2}) \) represents the binding interaction. 

Alternatively, it can be expressed in terms of the relative (\( r \)) 
and centre of mass coordinate (\( R \)):

\begin{eqnarray}
\hat{H} & = & \hat{H}_{r}+\hat{H}_{R}+v_{1}(R+\alpha _{2}r)+v_{2}(R-\alpha _{1}r),\label{eq:HrR} \\
\hat{H}_{r} & = & \frac{\hat{p}_{r}^{2}}{2\mu }+v(r)\label{eq:Hint} \\
\hat{H}_{R} & = & \frac{\hat{P}^{2}_{R}}{2M}\label{eq:HR} 
\end{eqnarray}
where \( M \) is the total mass, \( \alpha _{i}=m_{i}/M \) (\( i=1,2 \)),
and \( \hat{H}_{r} \) is the internal Hamiltonian. Here, \( \mu  \) represents
the reduced mass of the two-particle system and \( r=r_{1}-r_{2} \) is the
relative coordinate. 

We refer to the specific case in which \( v_{i}(r_{i}) \) corresponds to an
infinite wall potential, i.e.:

\begin{equation}
v_{i}(r_{i})=\left\{ 
\begin{array}{cc}
0 & r_{i}>0 \\
\infty  & r_{i}\leq 0
\end{array}\right. 
\end{equation}

For negative values of \( r_{1} \) and \( r_{2} \) the total wave function
must vanish. For positive values, the Hamiltonian comprises two terms: 
one associated
with the internal motion of the projectile (\( \hat{H}_{r} \)) and the other
describing the centre of mass motion (\( \hat{H}_{R}) \). 

We take \( v(r) \) to be a harmonic interaction. Therefore, the eigenfunctions
for the internal Hamiltonian are given by
\begin{equation}
\label{wfosc}
\phi _{n}(r)={\mathcal{N}}_{n}^{-\frac{1}{2}}\exp 
\left( -\frac{r^{2}}{2a^{2}_{0}}\right) {\mathcal H}_{n}\left( 
\frac{r}{a_{0}}\right) \, \, ;\, \, \, n=0,1,\ldots 
\end{equation}
 where \( a_{0}=\sqrt{\hbar /\mu \omega } \) is the oscillator 
length, \( {\mathcal{N}}_{n} \)
a normalisation constant and \( {\mathcal{H}}_{n} \) the Hermite polynomial of
order \( n \).

Denoting by \( N \) the number of open channels, the total wave function will
be then expanded in terms of eigenstates of the Hamiltonian as follows:

\begin{equation}
\label{eq:wf}
\Psi (r,R)=\frac{1}{\sqrt{v_{0}}}\phi _{0}(r)e^{-iK_{0}R}-
\sum ^{N-1}_{m=0}\frac{S_{0m}}{\sqrt{v_{m}}}\phi _{m}(r)e^{iK_{m}R}-
\sum ^{\infty }_{m=N}F_{0m}\phi _{m}(r)e^{-|K_{m}|R}.
\end{equation}
 
In this expression \( K_{m} \) represents the centre of mass 
momentum associated
with the internal state \( m \) and \( v_{m}=\hbar K_{m}/M \) its velocity.
Energy conservation applied to the whole system leads to the constrain

\begin{equation}
\label{eq:E(Kn,en)}
\frac{(\hbar K_{m})^{2}}{2M}+\epsilon _{m}=E,
\end{equation}
where \( \epsilon _{m} \) is the eigenvalue corresponding to the internal
state \( \phi _{m}(r) \).

The first term in (\ref{eq:wf}) represents an incoming wave, normalised to
unit flux, coming from \( R=+\infty  \). The second term contains the set of
scattered waves travelling in the positive \( R \) direction corresponding
to open channels, i.e., \( \epsilon _{m}<E \). The coefficients \( S_{nm} \)
are the S-matrix elements or, strictly speaking, reflection coefficients. Then,
\( S_{0m} \) represents the amplitude probability of populating the 
state \emph{\( m \)}
during the collision starting from a wave function in its ground state. The
last term in (\ref{eq:wf}) contains the contribution to the wave function due
to the (infinite) set of closed channels, for which \( \epsilon _{m}>E \).
For these states, the associated momentum \( K_{m} \) is a pure 
imaginary quantity,
and the centre of mass motion is described by an exponential 
decaying behaviour.
Therefore, they do not contribute to the asymptotic wave function and so they
do not give direct contribution to the outgoing flux. Although they are usually
ignored in practical calculations, the peculiarity of the infinite potential
requires however the inclusion of these states in order to describe correctly
the wave function in all the space.

\begin{figure*}
{\par\centering \resizebox*{0.75\textwidth}{!}{\rotatebox{-90}
{\includegraphics{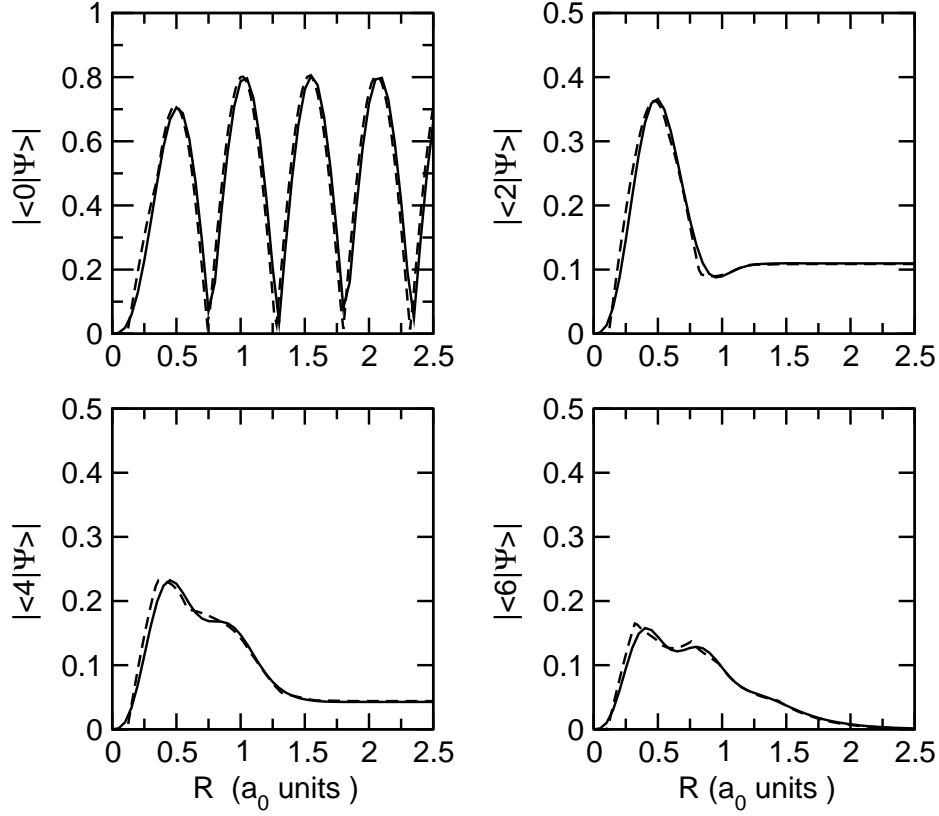}}} \par}
\caption{\label{Fig:WFmod_e5}Projection of the total wave functions on 
the internal
eigenstates \protect\( n\protect \)=0, 2, 4 and 6, for the scattering energy
\protect\( E=5\, \hbar \omega \protect \). The solid line corresponds to the
exact calculation by direct application of the boundary conditions, whereas
the dashed line corresponds to the exact calculation in the CLS approach. }
\end{figure*}

In this problem there is no transmitted wave due to the presence of the wall
and so the S-matrix should fulfil unitarity. Therefore, the ingoing
and outgoing flux must be equal:

\begin{equation}
\sum ^{N-1}_{m=0}|S_{nm}|^{2}=1,
\end{equation}
 where the sum extends to the set of open channels. 

The coefficients \( S_{nm} \) and \( F_{nm} \) are determined by imposing
the boundary conditions at \( r_{i}=0 \). For the infinite wall the wave function
must vanish at \( r_{1}=0 \) and \( r_{2}=0 \) which, in terms of \( r \)
and \( R \) means

\begin{eqnarray}
\Psi (r,R=+\alpha _{1}r)= & 0 & \,\,\,\,\,(r>0),\nonumber \\
\Psi (r,R=-\alpha _{2}r)= & 0 & \,\,\,\,\,(r<0).\label{eq:boundary_cond_wall} 
\end{eqnarray}

In what follows we present two alternative methods to solve \emph{exactly} this
problem.

\subsection{\label{sec:exact_direct}Exact solution by direct application 
of the boundary conditions}

As described above the exact solution of the scattering problem of a two-body
projectile by a rigid wall is accomplished by applying the boundary conditions
(\ref{eq:boundary_cond_wall}) to the general solution (\ref{eq:wf}). For 
simplicity we assume that the system is initially in its ground state. 
The first
condition in (\ref{eq:boundary_cond_wall}) leads to the following equation
for the scattering coefficients:
\begin{equation}
\label{eq:Phi(r,r/2)}
\frac{1}{\sqrt{v_{0}}}e^{-iK_{0}\alpha _{1}r}\phi _{0}(r)-\sum ^{N-1}_{n=0}
\frac{S_{0n}}{\sqrt{v_{n}}}e^{iK_{n}\alpha _{1}r}\phi _{n}(r)-
\sum ^{\infty }_{n=N}{F_{0n}}e^{-|K_{n}|\alpha _{1}r}\phi _{n}(r)=0\, ;\, r>0.
\end{equation}
 The second condition in (\ref{eq:boundary_cond_wall}) leads to a similar 
equation
(with \( \alpha _{2} \) instead of \( \alpha _{1} \)) which holds 
for \( r<0 \).
In particular, we have performed calculations for the particular case of equal
masses, i.e., \( \alpha _{1}=\alpha _{2}=\frac{1}{2} \), for which both 
equations
are identical, due to the symmetry of the problem under the exchange 
of \( r_{1} \)
and \( r_{2} \). In the remaining of this section we restrict to this 
particular case.

In order to transform eq. (\ref{eq:Phi(r,r/2)}) into an ordinary algebraic
equation, we require that 
\begin{equation}
\label{eq:integral_modulo_Phi}
\int _{0}^{+\infty }\left| \Psi (r,R=\frac{1}{2}r)\right| ^{2}dr=0,
\end{equation}
 which gives for the scattering coefficients the relation:

\begin{equation}
\label{eq:bound_1}
\frac{1}{2}-\sum ^{\infty }_{n}\left( C_{0n}a_{0n}+C^{*}_{0n}a_{0n}^{*}\right)
 +\sum ^{\infty }_{n,m}C_{0n}C^{*}_{0m}b_{nm}=0,
\end{equation}
 where

\begin{equation}
C_{0n}=\left\{ 
\begin{array}{cc}
\sqrt{\frac{v_{0}}{v_{n}}}S_{0n}\, \, ; & n<N \\
\sqrt{v_{0}}F_{0n}\, \, ; & n\geq N
\end{array}\right. 
\end{equation}
 and

\begin{eqnarray}
a_{0n} & = & \int _{0}^{\infty }\phi _{0}^{*}(r)\phi _{n}(r)e^{i(K_{n}+K_{0})
\frac{1}{2}r}\, \, ;\, \, \, n<N,\label{eq:aon} \\
b_{nm} & = & \int _{0}^{\infty }\phi _{n}^{*}(r)\phi _{m}(r)e^{i(K_{m}-K_{n})
\frac{1}{2}r}\, \, ;\, \, \, n,m<N.\label{eq:bnm} 
\end{eqnarray}
 
The expressions for \( a_{0n} \) and \( b_{nm} \) for \( n,m\geq N \) are
obtained substituting \( \pm iK_{n} \) for \( -|K_{n}| \). 

Differentiating eq. (\ref{eq:bound_1}) with respect to \( C^{*}_{0n} \) we
get the following linear system in the variables \( C_{0n} \):

\begin{equation}
\label{eq:Snm_direct}
\sum ^{\infty }_{m=0}C_{0m}b_{nm}=a^{*}_{0n},\, \, \, \, \, \, n=0,
\ldots ,\infty .
\end{equation}

For practical calculations the resolution of this system requires 
the truncation
of the sum at some finite value of \( m \). It should be noticed that in order
to achieve convergence for the S-matrix elements one is forced to include in
the calculation several closed channels. Otherwise, the boundary
conditions (\ref{eq:boundary_cond_wall}) are not accurately fulfilled.

\subsection{Exact solution using a discrete basis\label{sec:exact_CLS}}

In this section we present an alternative method to obtain the exact solution
of the problem stated above. The method relies on the introduction of a new
basis of states which are particular linear combinations of the internal wave
functions. The new functions have the property of being highly localised in
configuration space. As we shall see this peculiarity allows to apply 
more easily the boundary conditions. 

We start with a truncated basis of \( N \) eigenstates for the internal 
Hamiltonian
that we denote by \( \{|Nn\rangle ;n=1,\ldots ,N-1\} \). Thus, according to
our previous notation, \( \langle r|Nn\rangle =\phi _{n}(r) \). 

In the appendix we show how these states can be decomposed in terms 
of configuration localised states (CLS) as 

\begin{equation}
\label{eq:CLS}
|Nn\rangle =\sum ^{N}_{s=1}\left\langle CLS;Ns|Nn\right\rangle |CLS;Ns\rangle 
\end{equation}

In this expression the ket \( |CLS;Ns\rangle  \) represents a configuration
localised state. Explicit expressions for the CLS associated with the HO wave
functions can be found in the appendix \emph{}and in Ref.~\cite{Per99}. The
function \( \langle r|CLS;Ns\rangle  \) has the property of being highly 
localised around \( r=r_{s} \), the \( s \)-th zero of the eigenfunction 
\( \langle r|N\, N\rangle  \). 

The problem is significantly simplified in the case of fragments of 
equal masses.
Parity conservation guarantees that only states with the same parity as the
incident wave function will be populated in the process. This allows to work
in the subspace: \{\( \phi _{n}(r) \); \( n=0,2,\ldots ,N-2 \)\} 
(for simplicity
of the notation, and without loss of generality, we take \( N \) even). These
functions are symmetric with respect to their natural variable, \( r \). From
this set of \( \frac{N}{2} \) states it is possible to construct, by means
of a transformation similar to (\ref{eq:CLS}), a set of \emph{symmetric 
configuration
localised states} (SCLS), which are also even functions with respect to the
variable \( r \). We denote this new set of states by 
\{\( |SCLS;Ns\rangle ;\, s=1,\ldots ,\frac{N}{2} \)\},
where the index \( s \) runs over the positive zeros of 
\( {\mathcal{H}}_{N}(x) \).
The details of its derivation are presented in the appendix. The state 
\( \langle r|SCLS;Ns\rangle  \)
has the property of being localised around the points 
\( r=r_{s}=\pm x_{s}a_{0} \),
where \( x_{s} \) is the \( s- \)th positive zero of the Hermite polynomial
\( {\mathcal{H}}_{N}(x) \). In the treatment that follows we make extensive use
of this remarkable signature. 

The boundary condition due to the wall requires that the total wave function
vanishes for \( r_{1}=0 \) and \( r_{2}=0 \) or, in terms of coordinates 
\( r \)
and \( R \), along the lines \( R=r/2 \) (\( r>0 \)) and 
\( R=-r/2 \) (\( r<0) \)
in the (\( R,\, r \)) plane. At each value of \( R \) a reflected wave is
generated, interfering with the other outgoing waves to construct the total
scattered wave. This picture is simplified working in terms of SCLS. The part
of the wave function associated to the state \( \langle r|SCLS;Ns\rangle  \)
is peaked around \( r=\pm r_{s} \) and, therefore, it will be mainly scattered
around \( R=R_{s}=|r_{s}|/2 \). In the limit case \( N\rightarrow \infty  \),
\( \langle r|SCLS;Ns\rangle  \) becomes a delta function in \( r \) and the
associated wave is exactly scattered at \( R=R_{s} \). Moreover, continuity
of the wave function implies that a reflected wave, affected by a phase factor
\( -\exp (-2iK_{o}R_{s}) \), is generated at this point. Obviously, this is
not exactly our situation as, in practise, \( N \) is always finite and so
our localised states have a finite dispersion around \( r_{s} \). However,
we can make this dispersion as small as required by increasing the number of
states.

Consequently, the process is considered as a distribution of localised states
that are reflected at some definite \emph{barriers} in the \( R \) direction.
The total scattered wave is given by the superposition of the reflected waves.
To make the treatment clearer, we divide the \( R \) axis in \( \frac{N}{2} \)
regions, delimited by the values of \( R_{s} \). We order the zeros of the
Hermite polynomial \( {\mathcal{H}}_{N} \) in such way that 
\( R_{1}>R_{2}>\ldots >R_{N/2} \).
Let us introduce an index \( i \) to label each region 
(\( i=1,\ldots ,\frac{N}{2}) \),
such that \( i=1 \) corresponds to the asymptotic region, i.e., \( R>R_{1} \),
before any localised state has been reflected. In this region the basis space
associated with the internal motion is described in terms of the first 
\( \frac{N}{2} \)
HO eigenfunctions with positive parity. Alternatively, it can be described in
terms of \( \frac{N}{2} \) symmetric localised states. 

At \( R=R_{1} \) the SCLS corresponding to \( s=1 \) is reflected and removed
from the incident wave function, while the rest of SCLS remain unaltered. 
Therefore,
in the region \( i=2 \) the basis is limited to the subspace spanned by the
remaining \( \frac{N}{2}-1 \) states (\( s=2,...,\frac{N}{2} \)). Subsequently,
our original set of states are no longer eigenstates of the Hamiltonian in this
region. Instead, a new family of \( \frac{N}{2}-1 \) eigenstates must 
be calculated,
by diagonalizing the Hamiltonian in the basis constituted by the remaining 
\( \frac{N}{2}-1 \) localised states. 

The method is schematically illustrated in Fig.\ \ref{Fig:exact_cls}. The picture
represents the scattering problem in the \( (r,R) \) plane. The total wave
function travels in the \( R \) direction (vertical axis) and must vanish along
the lines \( r_{1}=0 \) and \( r_{2}=0 \), which have been also plotted for
reference. The case with \( N=8 \) has been considered, in which the incoming
wave is decomposed in a set of four SCLS. Each one of these SCLS is consider
to scatter at a definite barrier in the \( R \) axis, labelled from \( R_{1} \)
to \( R_{4} \). Note that they partially extend to the forbidden region 
(\( r_{i}<0 \)),
which is a consequence of the truncation of the HO basis.

\begin{figure}
{\par\centering \resizebox*{0.65\textwidth}{!}{\includegraphics{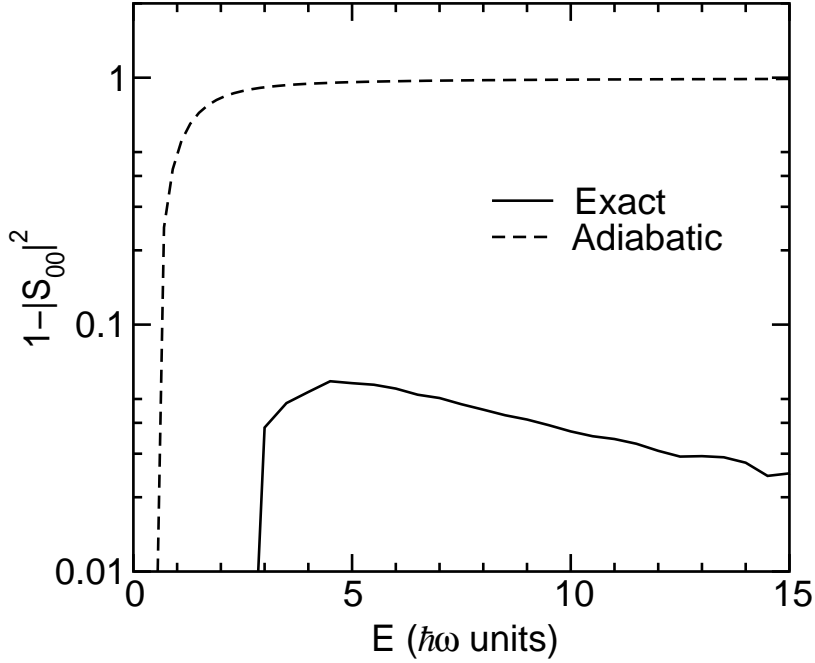}} \par}
\caption{\label{Fig:Sad}Representation of the quantity \protect\( 
1-|S_{00}|^{2}\protect \)
versus the scattering energy for the exact calculation (solid line) and the
adiabatic model (dashed line).}
\end{figure}

\begin{figure*}
{\par\centering \resizebox*{0.75\textwidth}{!}{\includegraphics{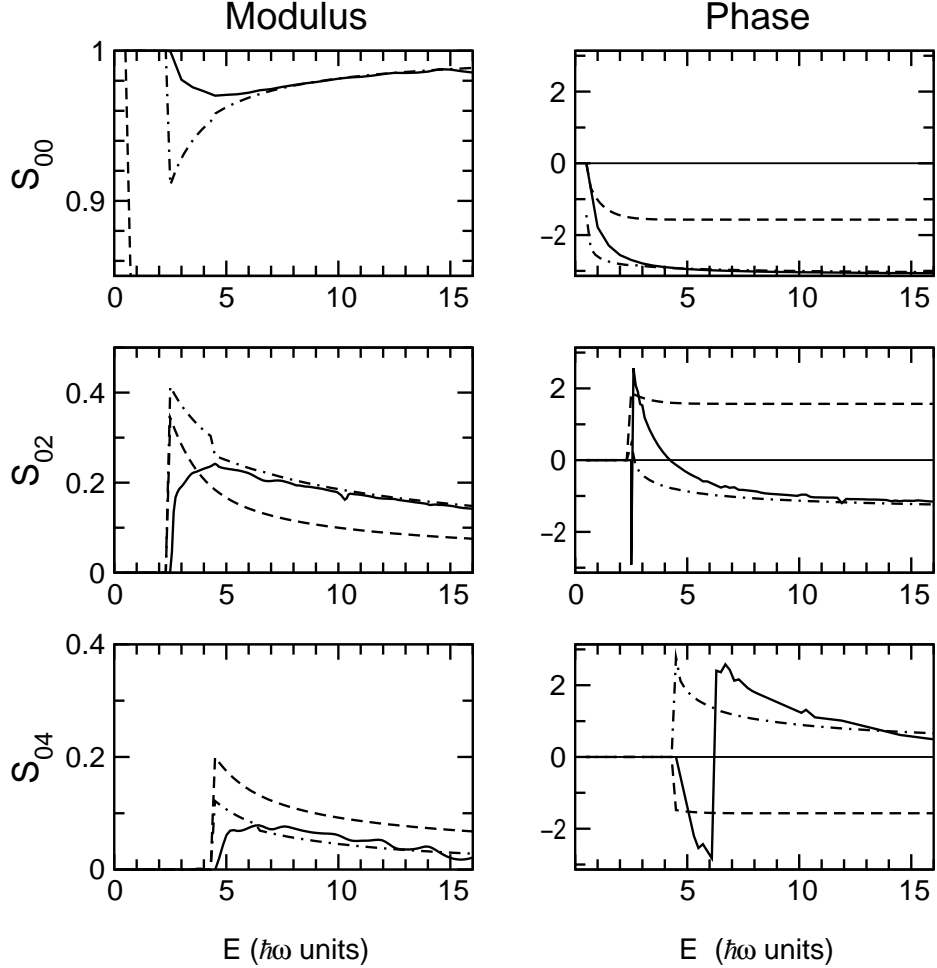}}
\par}
\caption{\label{Fig:Smat_usa}S-matrix coefficients for the ground and 
first excited
states. The solid line is the exact calculation. The dot-dashed line 
corresponds
to the USA model, with the prescription (\ref{eq:delta_ns}) for the phases.
The dashed line corresponds to the adiabatic model. }
\end{figure*}

Let us specify explicitly the boundary condition at \( R=R_{i} \). This barrier
separates regions \( i \) and \( i+1 \). The eigenstates corresponding to
region \( i \) will be denoted by \{\( |N^{(i)}\, m\rangle  \); 
\( m=0,...,N^{(i)}-1 \)\}
where \( N^{(i)}=\frac{N}{2}-i+1 \) is the number of states in region \( i \).
The total wave function is then expanded in each region in terms of the 
corresponding
eigenstates. For region \( i \), corresponding to the interval 
\( R_{i-1}>R>R_{i} \), we have:

\begin{equation}
\label{eq:Phi_CLS}
|\Psi ^{(i)}(E)\rangle =\sum ^{N^{(i)}-1}_{m=0}\left( A^{(i)}_{m}
e^{-iK^{(i)}_{m}R}-B^{(i)}_{m}e^{+iK^{(i)}_{m}R}\right) |N^{(i)}\, m\rangle ,
\end{equation}
 where \( A^{(i)}_{m} \) (\( B^{(i)}_{m} \)) are the coefficients 
of the incoming
(outgoing) waves. Energy conservation requires, in analogy with 
(\ref{eq:E(Kn,en)}),

\begin{equation}
\label{eq:E(Km,em)}
\frac{(\hbar K^{(i)}_{m})^{2}}{2M}+\epsilon ^{(i)}_{m}=E,
\end{equation}
 where \( \epsilon ^{(i)}_{m} \) is the \( m \)-th eigenvalue of the 
Hamiltonian
in region \( i \) and \( K^{(i)}_{m} \) its corresponding centre of mass 
momentum. 

Similarly, for region \( i+1 \) the wave function will be written as

\begin{equation}
|\Psi ^{(i+1)}(E)\rangle =\sum ^{N^{(i+1)}-1}_{n=0}\left( A^{(i+1)}_{n}
e^{-iK^{(i+1)}_{n}R}-B^{(i+1)}_{n}e^{+iK^{(i+1)}_{n}R}\right) |N^{(i+1)}\, 
n\rangle .
\end{equation}

At \( R=R_{i} \) the part of the wave function associated with the localised
state \( s=i \) is reflected by the wall. Therefore, continuity of the wave
function requires that

\begin{equation}
\label{eq:cont_Ri+1}
\left. \langle SCLS;N\, i|\Psi ^{(i)}(E)\rangle \right| _{R_{i}}=0.
\end{equation}

By contrast, the rest of localised states are unaffected by this barrier, so
we may require continuity of the wave function and its derivative for the 
components \( s>i \):

\begin{eqnarray}
\left. \langle SCLS;N\, s|\Psi ^{(i)}(E)\rangle \right| _{R=R_{i}} & = & 
\left. \langle SCLS;N\, s|\Psi ^{(i+1)}(E)\rangle \right| _{R=R_{i}}\\
\left[ \frac{d}{dR}\langle SCLS;N\, s|\Psi ^{(i)}(E)\rangle \right] _{R=R_{i}} & = & \left[ \frac{d}{dR}\langle SCLS;N\, s|\Psi ^{(i+1)}(E)\rangle 
\right]_{R=R_{i}}.\label{eq:condicion_Ri+1_b} 
\end{eqnarray}

Equations (\ref{eq:cont_Ri+1}) to (\ref{eq:condicion_Ri+1_b}) give rise to
a set of \emph{\( 2\left( \frac{N}{2}-i\right) +1 \)} conditions for the 
scattering
coefficients \( A^{(i)}_{n} \) and \( B^{(i)}_{n} \) 
(\emph{\( n=0,\ldots ,N^{(i)}-1 \)}).
When applied to all regions a total of \( \left( \frac{N}{2}\right) ^{2} \)
equations with \( \frac{N}{2}(\frac{N}{2}+1) \) coefficients is obtained. From
these, the \( \frac{N}{2} \) coefficients \( A^{(0)}_{n} \) are known, as
they are determined by the initial conditions. Therefore, there remain 
\( \frac{N}{2}(\frac{N}{2}+1)-\frac{N}{2}=\left( \frac{N}{2}\right) ^{2} \)
coefficients to be determined, that coincides with the number of equations.
Thus, the resolution of the problem reduces to the calculation of the inverse
of a \( \left( \frac{N}{2}\right) ^{2}\times \left( \frac{N}{2}\right) ^{2} \)
matrix. This can be a slow computational task when \( N \) takes large values.
This drawback has led us to adopt an alternative method that avoids this sort
of calculations, speeding significantly the computational time. 
For its application,
it is convenient to define a generalised ``S-matrix'' for region \( i \)
as:

\begin{equation}
\label{eq:BSA}
B^{(i)}_{n}=\sum _{n'}\sqrt{\frac{K_{n'}^{(i)}}{K^{(i)}_{n}}}S^{(i)}_{n'n}
A^{(i)}_{n'}.
\end{equation}

For region \( i=1 \) this definition gives just the usual S-matrix. The factor
\( \sqrt{K^{(i)}_{n'}/K^{(i)}_{n}} \) has been introduced to ensure unitarity. 

Substituting this expression into equations (\ref{eq:cont_Ri+1}) 
to (\ref{eq:condicion_Ri+1_b})
we get a set of \( 2\left( \frac{N}{2}-i\right) +1 \) equations relating the
S-matrix elements of \( S^{(i)} \) and \( S^{(i+1)} \). Then, instead of 
solving
the system of equations as a whole, we perform an iterative calculation 
in which
the S-matrix for each region is determined in terms of the S-matrix for the
neighbouring region. The starting point is the region \( i=\frac{N}{2} \),
for which the S-matrix is known. In this region there is only one state left,
which is completely reflected at \( R_{N/2} \). The S-matrix in this case is
just the phase factor: \( S^{(\frac{N}{2})}=\exp 
\big {(}-2iK^{(\frac{N}{2})}_{1}R_{N/2}\big {)} \).
From this, we can determine the S-matrix for region \( \frac{N}{2}-1 \), and
so on. Finally, we obtain the S-matrix, \( S^{(1)} \), in terms of 
\( S^{(2)} \).
Taking into account (\ref{eq:BSA}) and the fact that \( A^{(1)}_{n} \) are
given by the initial conditions, it is possible to determine the coefficients
\( B^{(1)}_{n} \). One can derive also an iterative procedure to determine
the coefficients \( A^{(i)}_{n} \) for all regions in terms of the 
corresponding S-matrices.

\subsection{Discussion of the exact results}

In order to compare both treatments we have plotted in 
figure \ref{Fig:Sexact(E)} the
S-matrix coefficients \( S_{00} \), \( S_{02} \) and \( S_{04} \), as 
a function
of the scattering energy. The energy scale is in units of \( \hbar \omega  \).
A basis space with 15 even parity HO states was used for the calculation. The
figure shows a good agreement between both treatments, specially at low 
scattering
energies. As the scattering energy increases, the effect of the truncation of
the basis becomes more important and both calculations differ slightly. Notice
that this discrepancy appears to be more evident for the phase of the S-matrix
elements. Nevertheless, this difference between both calculations is reduced
as the number of basis states is increased. 

We notice that, as expected, the elastic coefficient is identically one for
scattering energies below \( \frac{5}{2}\hbar \omega  \). This corresponds
to the energy of the first excited state that is suitable to be populated. 
Accordingly,
the inelastic coefficients, \( S_{02}, \) \( S_{04},\ldots  \) are identically
zero below this threshold. For scattering energies higher 
than \( \frac{5}{2}\hbar \omega  \),
the modulus of the elastic coefficient is less than one and the rest of the
coefficients are non zero. Note that the modulus of \( S_{00} \) presents a
minimum at \( E=5\, \hbar \omega  \), indicating a maximal loss of flux from
the elastic channel to other channels at this energy. For higher scattering
energies the modulus of elastic scattering coefficient tends gradually to 
unity.


We have also compared the wave functions in both approaches. In our problem
the wave function is a complex quantity depending on two variables, \( r \)
and \( R \). For the exact direct calculation the wave function is given by
(\ref{eq:wf}), where the scattering coefficients are calculated by imposing
the boundary conditions (\ref{eq:boundary_cond_wall}). It must be noticed,
however, that this expression only holds for \( r_{1},\, r_{2}\geq 0 \) or,
equivalently, for \( -2R<r<2R \). Outside this range, the total wave function
must be identically zero. The projection of the wave function on a state 
\( |N\, n\rangle  \) is calculated as

\begin{equation}
\label{eq:<n|Phi>}
\Psi _{n}(R)\equiv \langle N\, n|\Psi \rangle =\int ^{2R}_{-2R}dr
\phi ^{*}_{n}(r)\Psi (r,R).
\end{equation}

For large values of \( R \) (compared to the spatial extension of the internal
wave function) the integral can be extended to the interval 
\( (-\infty ,\infty ) \)
and the projection above is directly related to the corresponding 
S-matrix element:

\begin{equation}
\label{eq:<n|Phi>_asym}
\Psi _{n}(R)\approx \int ^{\infty }_{-\infty }dr\phi ^{*}_{n}(r)\Psi (r,R)=
\frac{1}{\sqrt{v_{0}}}\delta _{n0}e^{-iK_{0}R}-\frac{1}{\sqrt{v_{n}}}S_{0n}
e^{iK_{n}R}\, ;\, \, \, (R\gg a_{0}).
\end{equation}

In the analysis based on CLS, the total wave function is given in terms of a
piecewise function of \( R \), according to (\ref{eq:Phi_CLS}). For a certain
region \( i \), this wave function is written as a superposition of the 
eigenstates
of the Hamiltonian in this region. These eigenstates travel freely between the
consecutive barriers \( R_{i} \) and \( R_{i+1} \). Thus, the projection on
a state \( |Nn\rangle  \) for a value of \( R \) belonging to a region \( i \)
reads

\begin{equation}
\Psi ^{(i)}_{n}(R)=\langle N\, n|\Psi ^{(i)}\rangle =\sum ^{N^{(i)}-1}_{m=0}
\left( A^{(i)}_{m}e^{-iK^{(i)}_{m}R}-B^{(i)}_{m}e^{+iK^{(i)}_{m}R}\right) 
\langle N\, n|N^{(i)}\, m\rangle .
\end{equation}

We notice that, except for the incoming region (\( i=1 \)), the states 
\( |N\, n\rangle  \)
and \( |N^{(i)}\, m\rangle  \) are eigenstates belonging to different 
Hamiltonians
and so orthogonality can not be directly applied to them. However, they are
both given in terms of the SCLS and so the calculation of their overlap is 
straightforward.

The results are shown in figure \ref{Fig:WFmod_e5}, where we present the 
modulus
of the projection of the total wave function on the internal states. The solid
line corresponds to the treatment of section~\ref{sec:exact_direct} and the
dashed line to the CLS method. These wave functions have been obtained for a
scattering energy \( E=5\, \hbar \omega  \). The \( R \) scale is in units
of \( a_{0} \). As shown in this figure both treatments give almost identical
results, the small differences being attributed to the truncation of the 
infinite basis. 

A similar agreement between both approaches is found at other energies, 
provided that a sufficient number of states is included in each case. 

From this picture it becomes apparent the role played by each internal state.
The state \( n \)=0 corresponds to the initial state. Then, the asymptotic
wave function contains both incoming and outgoing contributions of this state.
This is reflected in the characteristic interference pattern of the curve 
\( |\langle N\, 0|\Psi \rangle | \).
At this scattering energy, the states \( n \)=2 and \( n \)=4 are open 
channels,
and so they give a non vanishing contribution to the asymptotic scattered wave.
In fact, the curves tend to a constant value for large distances which, 
according
(\ref{eq:<n|Phi>_asym}), is proportional to the corresponding coefficient 
\( S_{0n} \).
By contrast, the state \( n \)=6 is closed. Its exponential
decaying tail reflects the fact that this state does not 
contribute \emph{directly}
to the asymptotic wave function, but it does give a non negligible contribution
to the total wave function in the vicinity of the wall. As mentioned before,
the inclusion of these states is essential in order to reproduce accurately
the boundary conditions and to achieve convergence for the S-matrix elements.
Then, they indirectly affect the wave function in all the space.

The approach based on CLS presents some advantages compared to the treatment
described in section~\ref{sec:exact_direct}. It allows to evaluate all the
matrix elements \( S_{nm} \), \( n,m=0,\ldots ,N-1 \) in a single calculation,
i.e., the initial state does not need to be specified. By contrast, in 
the previous
approach a new calculation is required for each initial state. 

Moreover, the method based on the CLS preserves the general properties of the
S-matrix for any value of \( N \). In particular, conditions 
(\ref{eq:cont_Ri+1})
and (\ref{eq:condicion_Ri+1_b}) ensure that the total flux is conserved at
each barrier. As a relevant consequence, the resulting S-matrix fulfils 
unitarity,
regardless of the number of initial states chosen for the basis.

\section{The adiabatic model\label{sec:adiab}}

We derive in this section an expression for the S-matrix in the adiabatic 
approximation
\cite{Ron70}. The standpoint of the approach is that a fundamental distinction
is made between the two relevant coordinates of our problem, namely, the centre
of mass coordinate, \( R \), and the internal variable, \( r \). The former
is identified as a high-energy (\emph{fast}) variable and the latter 
as a low-energy (\emph{slow)} variable. 

In this dynamical regime, it is expected that the excitation 
energies \( \epsilon _{n} \)
associated with those excited states which are significantly populated, are
such that \( \epsilon _{n}\ll E \), where \( E \) is the incident energy of
the projectile. Under this assumption it seems reasonable to replace the 
internal
Hamiltonian by a representative constant. By choosing this constant 
as \( \epsilon _{0} \),
the ground state energy, it is also guaranteed that the solution of 
the resulting
approximate three-body equation satisfies the correct incident wave boundary
condition. 

Then, applying this approximation to the Hamiltonian (\ref{eq:HrR}) 
the Schr\"odinger equation reads:

\begin{equation}
\label{eq:Sch_ad}
\left[ -\frac{\hbar ^{2}}{2M}\frac{d^{2}}{dR^{2}}+v_{1}(R+\frac{1}{2}r)+
v_{2}(R-\frac{1}{2}r)+\epsilon _{0}-E\right] \Psi ^{ad}(R,r)=0.
\end{equation}

Note that this approximate Schr\"odinger equation is independent of the 
relative
momentum between the fragments. Then, its conjugate coordinate, \( r \), is
a constant of motion, remaining \emph{frozen} during the collision. Thus, eq.\
(\ref{eq:Sch_ad}) has to be solved for all values of a fixed separation 
\( r \).

In the case of a rigid wall potential, \( v_{1} \) and \( v_{2} \) are zero
for positive values of \( r_{1} \) and \( r_{2} \), respectively. Therefore,
in this case the solution of eq.\ (\ref{eq:Sch_ad}) is given by the plane wave,
\( \exp (-iK_{0}R) \), multiplied by an arbitrary function of \( r \). The
most general solution verifying the boundary incident condition at infinity
can be written as

\begin{equation}
\label{eq:ad_wf}
\Psi ^{ad}(r,R)=\phi _{0}(r)e^{-iK_{0}R}-S(r)\phi _{0}(r)e^{iK_{0}R}.
\end{equation}
 where \( \phi _{0}(r) \) is the ground state wave function and \( S(r) \)
is a function determined by imposing the boundary condition at the wall. This
requires that the wave function vanishes at \( R=|r|/2 \). Then 

\begin{equation}
\label{eq:S(r)}
S(r)=e^{-iK_{0}|r|}.
\end{equation}

The scattering coefficients defined by eq. (\ref{eq:wf}) can be obtained by
projecting the wave function (\ref{eq:ad_wf}) onto the basis states 
\( \{\phi _{n}(r)\} \).
This gives rise to the following simple expression:

\begin{equation}
\label{eq:Son_ad}
S_{0n}=\int ^{+\infty }_{-\infty }\phi _{n}^{*}(r)S(r)\phi _{0}(r)dr.
\end{equation}

We notice that in this expression no distinction is made between open
and closed channels. Actually, the approximation treats the full excitation
spectrum of the internal Hamiltonian as being degenerate in energy with the
ground state. As a consequence, unitarity of the adiabatic S-matrix is only
achieved when summing over the infinite set of eigenstates.

In what follows we show that the adiabatic approximation is not adequate for
the problem treated in this work. In Fig.\ \ref{Fig:Sad} the quantity 
\( 1-|S_{00}|^{2} \)
is plotted versus the collision energy for the exact (solid line) and adiabatic
(dashed line) calculations. This quantity can be interpreted as an excitation
probability. The adiabatic prediction completely disagrees with the 
exact calculation,
indicating that the assumptions involved in the adiabatic approximation are
not adequate for this problem. We attribute this failure to the fact that, as
revealed by the exact calculation, many internal states participate in this
process\emph{.}

This is probably due to the peculiarities of an infinite zero-range 
interaction.
In this case, the momentum transferred to each particle by the wall is twice
the incident momentum and so, when increasing the scattering energy, 
the expected
excitation energy increases. During the collision time, i.e., while 
one particle
has collided with the wall, but the other still has not, the internal motion
of the projectile is strongly excited. Then, the assumption of the adiabatic
model, i.e., to consider the whole spectrum to be degenerated with the ground
state, does not work properly in this case. After the collision, that is, when
both particles have collided with the wall, the centre of mass momentum is 
reversed
and the final excitation energy is small. Thus, this model represents a case
in which, although the final excitation of the projectile is small, the 
adiabatic
approximation is inadequate, because during the collision the internal motion
is strongly excited.

From this discussion we conclude that one should be very careful in applying
the adiabatic approximation when dealing with strong, short-range interactions.
In the next section we develop a new method to treat this type of situations.

\section{Uncorrelated Scattering Approximation (USA)\label{sec:our_model}}

The main goal of this section is to derive an approximated expression for the
S-matrix corresponding to the one dimensional scattering of a two particle 
system
in terms of the constituent-target scattering amplitudes. The results presented
here are not restricted to the case of an infinite potential. Thus, we start
with a general derivation of the model ant later we particularize the results
to the problem of a wall potential in order to compare with the exact 
solution. 

There are two opposite effects acting on a projectile in the process of the
collision. The first one is the binding potential \( v(r) \) that tends to
keep the system bound. The second one is the interaction with the target which,
apart from governing the motion of the projectile centre of mass, is the 
responsible
for exciting or breaking the system. The relative importance of these two 
effects
depends importantly on the separation between the projectile and the target.
In particular, for sufficient large distances between them the dominant 
interaction
is clearly the mutual interaction between the constituents. Thus, 
it seems reasonable
to approximate the projectile-target potential by an average (folding) 
potential
at sufficiently large distances. By contrast, when the  bound projectile
is close enough to the target the dynamical evolution of the projectile 
is mainly
governed by the target interaction. In this case it is reasonable to neglect
the correlations between the fragments. 

Let us introduce a characteristic centre of mass distance \( R_{0} \) 
separating
these two regions. For distances \( R\gg R_{0} \), referred as the ``asymptotic
region'' and denoted by the index I, we adopt an approximate Hamiltonian in
which the interaction with the target is neglected

\begin{equation}
\label{eq:HI}
\hat{H}\approx \hat{H}_{r}+\hat{H}_{R}\equiv \hat{H}_{I},
\end{equation}
 where \( \hat{H}_{r} \) is the internal Hamiltonian (\ref{eq:Hint}) and 

\begin{equation}
\hat{H}_{R}=\frac{{\hat{P}}^{2}_{R}}{2M}+V_{F}(R)
\end{equation}
with \( V_{F}(R) \) representing the folding potential between the projectile
and target. 

Asymptotically, the eigenstates of \( \hat{H}_{I} \) are just the product of
the eigenstates of the internal Hamiltonian \( \hat{H}_{r} \) times a plane
wave in \( R \), subject to the restriction (\ref{eq:E(Kn,en)}). As mentioned
before, there is only contribution to the asymptotic wave function coming from
the open channels. Denoting by \( N \) the number of these states, we restrict
the basis space to the set \( \{|Nn\rangle ;\, n=0,1,\ldots ,N-1\} \). Then,
the total wave function corresponding to an incoming wave in an internal state
\( n \) will be written in the asymptotic region as

\begin{equation}
\label{eq:wf_asym_USM}
|\Psi _{I,n}(E)\rangle \rightarrow \frac{1}{\sqrt{v_{n}}}e^{-iK_{n}R}|Nn
\rangle -\sum ^{N-1}_{m=0}\frac{S_{nm}}{\sqrt{v_{m}}}e^{iK_{m}R}|Nm\rangle .
\end{equation}

For distances \( R\ll R_{0} \), in what we call ``interaction region'' (denoted
by II), the Hamiltonian (\ref{eq:Hr1r2}) is approximated by

\begin{equation}
\label{eq:HII}
\hat{H}\approx \hat{h}_{1}+\hat{h}_{2}+\bar{v}\equiv \hat{H}_{II},
\end{equation}
 where \( \bar{v} \) is a constant that substitutes \( v(r) \).

The eigenfunctions of \( \hat{H}_{II} \) can be expanded in terms of the 
product
of eigenfunctions of the Hamiltonians \( \hat{h}_{1} \) and \( \hat{h}_{2} \).
An eigenstate of \( \hat{h}_{i} \) corresponds to the distorted wave for the
scattering of a particle under the potential \( v_{i}(r_{i}) \):

\begin{equation}
\hat{h}_{i}|\chi _{i}(k_{i})\rangle =E_{i}|\chi _{i}(k_{i})\rangle 
;\, \, \, \, \, (i=1,2),
\end{equation}
 where \( k_{i} \) is the asymptotic incident momentum for the constituent
\( i \), and \( E_{i}=(\hbar k_{i})^{2}/2m_{i} \). 
Asymptotically this distorted wave behaves as 

\begin{equation}
\label{eq:DistWaves}
|\chi _{i}(k_{i})\rangle \rightarrow |k_{i}\rangle -S_{i}(k_{i})|-k_{i}
\rangle ;\, \, \, \, \, (i=1,2),
\end{equation}
 where \( \langle r_{i}|k_{i}\rangle =\exp (-ik_{i}r_{i})/\sqrt{2\pi } \) is
a plane wave with momentum \( k_{i} \) and \( S_{i}(k_{i}) \) is 
the constituent-target
S-matrix for the scattering energy \( E_{i} \). Note that, in the case of the
infinite wall, this expression is valid for all values of \( r_{i}\geq 0 \).

Thus, the eigenstates of the Hamiltonian \( \hat{H}_{II} \), corresponding
to the asymptotic momenta \( k_{1} \) and \( k_{2} \), can be expressed as
products of the form

\begin{equation}
\label{eq:wf(k1,k2)}
|\psi (k_{1},k_{2})\rangle \propto |\chi _{1}(k_{1})\rangle 
|\chi _{2}(k_{2})\rangle .
\end{equation}
 where

\begin{equation}
\label{eq:E(e1,e2,v)}
\frac{\hbar ^{2}k^{2}_{1}}{2m_{1}}+\frac{\hbar ^{2}k^{2}_{2}}{2m_{2}}
+\bar{v}=E.
\end{equation}

According to (\ref{eq:DistWaves}), the asymptotic expansion of the 
eigenfunction
(\ref{eq:wf(k1,k2)}) contains an incoming wave, 
\( |k_{1}\rangle |k_{2}\rangle  \),
an scattered wave, \mbox{\( S_{1}(k_{1})S_{2}(k_{2})|-k_{1}\rangle 
|-k_{2}\rangle  \)},
and two cross terms mixing ingoing and outgoing contributions, 
namely \( S_{1}(k_{1})|-k_{1}\rangle |k_{2}\rangle  \)
and \( S_{2}(k_{2})|k_{1}\rangle |-k_{2}\rangle  \). In practise, these last
two terms do not contribute to the incoming nor the scattered wave function
at large distances. This can be verified by considering an incoming wave packet
in \( k_{1} \) and \( k_{2} \). The superposition of states of the form 
\( |k_{1}\rangle |-k_{2}\rangle  \)
or \( |-k_{1}\rangle |k_{2}\rangle  \) contains an incoming part which vanishes
for \( t\rightarrow +\infty  \), and an outgoing part which cancels for 
\( t\rightarrow -\infty  \).
Then, these cross terms can be omitted as far as the asymptotic 
behaviour concerns.
As we will show later, the use of MLS allows to demonstrate that these terms
do not contribute to the wave function at large distances in a time-independent
formalism. The form of the scattered wave, 
\( S_{1}(k_{1})S_{2}(k_{2})|-k_{1}\rangle |-k_{2}\rangle  \),
indicates that the S-matrix for an incoming wave with definite values of 
\( k_{1} \)
and \( k_{2} \), denoted \( S(k_{1},k_{2}) \), appears to be the product of
the individual S-matrices for the constituents, i.e., 
\( S(k_{1},k_{2})=S_{1}(k_{1})S(k_{2}) \).
This S-matrix is unitary, provided the individual S-matrices are unitary, i.e.,
\( |S_{i}(k_{i})|=1 \). This condition is satisfied for the infinite potential,
but it also holds for any other situation for which the transmission 
coefficient is zero.

Therefore, the scattering problem for the Hamiltonian 
\( \hat{H}_{II} \) corresponding
to a situation characterised by an incoming wave with definite values of the
energies of the constituents can be easily solved. However, our physical 
initial
state is not characterised by the individual energies of the two particles,
but by a certain internal state of the projectile and the energy of 
the collision.
The general solution in the interaction region for a total energy \( E \) will
be a certain superposition of eigenstates (\ref{eq:wf(k1,k2)}), verifying 
(\ref{eq:E(e1,e2,v)})
and the adequate asymptotic boundary conditions. These boundary 
conditions require
that the wave function in region II matches smoothly with the asymptotic 
wave function
of eq.\ (\ref{eq:wf_asym_USM}). One possible way to proceed might be to expand
the total wave function in each region in terms of the eigenstates of 
the approximated
Hamiltonian for that region. The coefficients of the expansion are determined
by imposing the continuity of the wave function and its derivative at 
the matching
radius \( R_{0} \). Apart from the complexity of the calculation, this method
has the problem that incoming waves coming from the asymptotic region do not
match exactly with incoming waves of the interaction region, due to 
the discontinuity
of the Hamiltonian at \( R_{0} \). As a result, part of the incoming flux is
reflected at \( R_{0} \) and spurious outgoing waves are generated. To overcome
this difficulty we proceed on a different way. In order to avoid the unphysical
reflections we relax the meaning of the matching radius \( R_{0} \). We assume
that the Hamiltonian \( \hat{H}_{I} \) is smoothly transformed into 
\( \hat{H}_{II} \)
in a finite transition region around \( R_{0} \). Although we do not make an
explicit description of this transition region in our model, we include its
effect by requiring that the wave function passes from one region to the other
without loss of flux. 

Let us consider the incoming part of the general solution 
(\ref{eq:wf_asym_USM}):

\begin{equation}
|\Psi ^{(in)}_{I,n}(E)\rangle \rightarrow \frac{1}{\sqrt{v_{n}}}e^{-iK_{n}R}
|Nn\rangle .
\end{equation}

In order to match this wave function with the inner wave solution it is 
convenient
to express the internal states in terms of a basis of Momentum Localised States
(MLS). These are obtained from the original basis by diagonalizing the momentum
operator in the set of internal states. Due to the truncation of the original
basis the MLS do not have a definite value of the internal momentum \( q \)
but, provided the number of states \( N \) is large, their momentum 
distribution
is highly localised around a certain value. The two basis of states 
are connected
by an orthogonal transformation (see appendix) and so, the incoming state can
be rewritten as

\begin{equation}
\label{eq:WFi_in}
|\Psi ^{(in)}_{I,n}(E)\rangle \rightarrow \frac{1}{\sqrt{v_{n}}}e^{-iK_{n}R}
\sum ^{N}_{s=1}\left\langle MLS;Ns|Nn\right\rangle |MLS;Ns\rangle .
\end{equation}
 where \( |MLS;Ns\rangle  \) denotes a MLS and \( \left\langle MLS;Ns|Nn
\right\rangle  \)
are the transformation coefficients. The function \( \langle q|MLS;Ns\rangle 
\equiv \tilde{\varphi }_{s}(q) \)
has the property of being highly localised around \( q=q_{s} \), the \( s \)-th
zero of the eigenfunction \( \langle q|N\, N\rangle  \).

As noted before, the eigenstates of the approximated Hamiltonian in the 
interaction
region are characterized by the energies of the two particles. These energies
are directly related to their incident momenta. Equivalently, they can 
be characterised
by the asymptotic values of the internal momentum \( q \) and the centre of
mass momentum \( K \). In our approach, the internal eigenstates will 
be approximated
in region II by the discrete basis of MLS. The incoming wave function 
for the interaction
region is then expressed at large distances as

\begin{equation}
\label{eq:WFin_II }
|\Psi ^{(in)}_{II,n}(E)\rangle \rightarrow \sum ^{N}_{s=1}A_{s}^{(n)}
e^{-i\tilde{K}_{s}R}|MLS;Ns\rangle 
\end{equation}
 with the centre of mass momentum \( \tilde{K}_{s} \) defined by the relation

\begin{equation}
\label{eq:E(qs,Ks,v)}
\frac{\hbar ^{2}q_{s}^{2}}{2\mu }+\frac{\hbar ^{2}{\tilde{K}}_{s}^{2}}{2M}
+\bar{v}=E.
\end{equation}

The coefficients \( A^{(n)}_{s} \) are determined in order the waves 
(\ref{eq:wf_asym_USM})
and (\ref{eq:WFin_II }) match smoothly. In particular, we impose the incoming
flux to be conserved in the transition. This can be achieved by taking

\begin{equation}
\label{eq:As}
A^{(n)}_{s}=\frac{1}{\sqrt{\tilde{v}_{s}}}e^{-i\delta _{ns}}\left\langle MLS;
Ns|Nn\right\rangle ,
\end{equation}
 with \( \tilde{v}_{s}=\hbar \tilde{K}_{s}/M \) and \( \delta _{ns}=
\gamma _{n}-\chi _{s} \).
The phases \( \gamma _{n} \) and \( \chi _{s} \) must be real numbers 
to preserve the incoming flux.

In order to evaluate the phases \( \gamma _{n} \) and \( \chi _{s} \), we
impose the two incoming solutions to have the same phase at \( R_{0} \). This
is achieved by taking \( \gamma _{n}=K_{n}R_{0} \) and \( \chi _{s}=
\tilde{K}_{s}R_{0} \),
which leads to

\begin{equation}
\label{eq:delta_ns}
\delta _{ns}=(K_{n}-\tilde{K}_{s})R_{0}.
\end{equation}

Once the coefficients \( A^{(n)}_{s} \) are known, the total wave function
in the interaction region is completely determined. It can be expressed 
in terms
of the distorted waves for each one of the constituents. To this end, 
we rewrite expression (\ref{eq:WFin_II }) as

\begin{eqnarray}
|\Psi ^{(in)}_{II,n}(E)\rangle  & \rightarrow  & \sum ^{N}_{s=1}A_{s}^{(n)}
e^{-i\tilde{K}_{s}R}\int _{-\infty }^{\infty }dq|q\rangle \tilde{\varphi }_{s}
(q)=\nonumber \\
 & = & \sum ^{N}_{s=1}A_{s}^{(n)}\int _{-\infty }^{\infty }dq\, 
\tilde{\varphi }_{s}(q)|k^{s}_{1}(q)\rangle |k^{s}_{2}(q)\rangle ,
\label{eq:WFii_in(k1,k2)} 
\end{eqnarray}
 where we have introduced the momenta \( k^{s}_{1}(q)=\frac{m_{1}}{M}
\tilde{K}_{s}+q \)
and \( k^{s}_{2}(q)=\frac{m_{2}}{M}\tilde{K}_{s}-q \). The scattering wave
function corresponding to an incoming plane wave \( |k^{s}_{i}(q)\rangle  \)
is given by the distorted wave \( |\chi (k^{s}_{i}(q)\rangle  \). Then, the
total wave function in region  II,  including both the incoming and scattered
wave reads

\begin{equation}
\label{eq:WFii_(X1,X2)}
|\Psi _{II,n}(E)\rangle =\sum ^{N}_{s=1}A_{s}^{(n)}\int _{-\infty }^{\infty }
dq\, \tilde{\varphi }_{s}(q)|\chi _{1}(k^{s}_{1}(q))\rangle |\chi _{2}
(k^{s}_{2}(q))\rangle .
\end{equation}

Taking into account the asymptotic behaviour of the distorted waves,
(\ref{eq:DistWaves}),
this wave function can be written beyond the range of the potentials as

\begin{eqnarray}
|\Psi _{II,n}(E)\rangle  & \rightarrow  & \sum ^{N}_{s=1}A_{s}^{(n)}
\int _{-\infty }^{\infty }dq\, \tilde{\varphi }_{s}(q)\Big {\{}|k^{s}_{1}(q)
\rangle |k^{s}_{2}(q)\rangle \nonumber \\
 & - & S_{1}(k^{s}_{1}(q)|-k^{s}_{1}(q)\rangle |k^{s}_{2}(q)\rangle 
-S_{2}(k^{s}_{2}(q)|k^{s}_{1}(q)\rangle |-k^{s}_{2}(q)\rangle \nonumber \\
 & + & S_{1}(k^{s}_{1}(q)S_{2}(k^{s}_{2}(q)|-k^{s}_{1}(q)\rangle 
|-k^{s}_{2}(q)\rangle \Big {\}}
\end{eqnarray}

Assuming that the individual S-matrices, \( S_{1} \) and \( S_{2} \), are
smooth functions of the energy in the region where the integrand takes
significant
values, they can be evaluated at \( k^{s}_{1} \) and \( k^{s}_{2} \), 
respectively.
Also, it is convenient to express the products of planes waves in terms of the
relative and centre of mass momenta. Considering the case of equal masses we
can use the expressions \( k^{s}_{1}(q)r_{1}+k^{s}_{2}(q)r_{2}
=\tilde{K}_{s}R+qr \)
and \( k^{s}_{1}(q)r_{1}-k^{s}_{2}(q)r_{2}=\frac{1}{2}\tilde{K}_{s}r+2qR \).
Then one can perform explicitly the integration with respect to \( q \), to
obtain

\begin{eqnarray}
\Psi _{II,n}(r,R) & \rightarrow  & \sum ^{N}_{s=1}A_{s}^{(n)}\Big {\{}
e^{-i\tilde{K}_{s}R}\varphi ^{*}_{s}(r)+S_{1}(k^{s}_{1})S_{2}(k^{s}_{2})
e^{i\tilde{K}_{s}R}\varphi _{s}(r)\nonumber \\
 & - & S_{1}(k^{s}_{1})e^{i\frac{1}{2}\tilde{K}_{s}r}\varphi _{s}(2R)-S_{2}
(k^{s}_{2})e^{-i\frac{1}{2}\tilde{K}_{s}r}\varphi ^{*}_{s}(2R)\Big {\}}
\nonumber \\
 & \label{eq:WFii(r,R)} 
\end{eqnarray}
 where \( \varphi _{s}(r) \) is the Fourier transform of 
\( \tilde{\varphi }_{s}(q) \).
This wave function can be interpreted as follows. The incoming wave 
is decomposed
as products of MLS, \( \varphi _{s}(r) \), describing the internal evolution,
times an incoming plane wave describing the centre of mass motion, 
\( e^{-i\tilde{K}_{s}R} \).
This wave scatters by the target giving rise to three terms. The second term
in (\ref{eq:WFii(r,R)}) is just the conjugate of the incident wave, times the
product of the S-matrices of the constituents. The remaining two terms comprise
the product of the function \( \varphi _{s}(2R) \), or its conjugate, times
a plane wave in the variable \( r \). As it can be easily verified, 
the function
\( \varphi _{s}(2R) \) vanishes for large values of \( R \) and so, these
two terms do not contribute to the asymptotic wave function. Actually, these
terms containing only one of the involved S-matrices can be physically regarded
as the situation in which only one of the particle has scattered and the other
has not yet. This is consistent with the fact that they both cancel at large
distances. However, we remark that these vanishing terms are essential in order
to reproduce the wave function at small distances. In this sense, it is also
interesting to note that the wave function (\ref{eq:WFii(r,R)}) 
retains components
associated to closed channels, even when they are explicitly omitted in the
asymptotic region, according (\ref{eq:wf_asym_USM}). As noted in 
section~II,
in the case of the wall potential the inclusion of these states is essential
in order to reproduce the boundary conditions. For this particular problem,
expression (\ref{eq:WFii(r,R)}) is valid for all the interaction region and,
as can be easily verified, it identically fulfils the boundary conditions, 
vanishing for \( r=\pm 2R \). 

Thus, we can conclude that an incoming MLS scatters with the product of the
S-matrices of the fragments. This result is consistent with our 
previous discussion
in which, using wave packet arguments, we concluded that the S-matrix is given
by the product of the S-matrices of the fragments in the basis characterised
by the momenta of the two particles. We now see that this property also holds
for the MLS basis which, in a sense, can be described as a wave packet of plane
waves in terms of \( q \), centred around \( q_{s} \).

Therefore, writing explicitly the value of the coefficients \( A^{(n)}_{s} \)
the scattered wave in the interaction region behaves at large distances as

\begin{equation}
\label{eq:WFscat_II}
|\Psi ^{(scat)}_{II,n}(E)\rangle \rightarrow \sum ^{N}_{s=1}\frac{1}
{\sqrt{\tilde{v}_{s}}}e^{-i\delta _{ns}}\left\langle MLS;Ns|Nn\right\rangle 
e^{i\tilde{K}_{s}R}S_{1}(k^{s}_{1})S_{2}(k^{s}_{2})|MLS;Ns\rangle .
\end{equation}

By writing the MLS appearing in this expression in terms of the original basis,
and imposing the conservation of flux, one can easily obtain the scattered wave
in the asymptotic region

\begin{eqnarray}
|\Psi ^{(scat)}_{I,n}(E)\rangle  & \rightarrow  & \sum _{m}\frac{e^{iK_{m}R}}
{\sqrt{v_{m}}}\Big {\{}\sum ^{N}_{s=1}e^{-i(\delta _{ns}+\delta _{ms})}
\left\langle Nm|MLS;Ns\right\rangle \nonumber \\
 & \times  & \langle MLS;Ns|Nn\rangle S_{1}(k^{s}_{1})S_{2}(k^{s}_{2})
\Big {\}}|Nm\rangle .
\end{eqnarray}

The expression between brackets provides the S-matrix element connecting an
initial state \( n \) with a final state \( m \)

\begin{equation}
\label{eq:derived_Snm}
S_{nm}=-\sum ^{N}_{s=1}e^{-i(\delta _{ns}+\delta _{ms})}\left
\langle Nm|MLS;Ns\right\rangle S_{1}(k^{s}_{1})S_{2}(k^{s}_{2})\left\langle
MLS;Ns|Nn\right\rangle .
\end{equation}

As pointed before, the peculiarities of the problem treated in this work imply
that, starting with the system in its ground state, only positive parity states
are suitable to be populated. As far as expression (\ref{eq:derived_Snm})
concerns,
this means that the matrix elements \( S_{0n} \), with \( n \) odd, 
are identically zero. 

Under these considerations it is possible to exclude those inhibited states.
Then, one is left with a truncated basis of the form \( |N\, n\rangle  \),
where \( N \) is now even. Following the arguments above and using the results
of the appendix one finally gets the following expression for the S-matrix in
the case of equal masses

\begin{equation}
S_{nm}=-\sum _{s}e^{-i(\delta _{ns}+\delta _{ms})}\left\langle N\, m|SMLS;N\, 
s\right\rangle \left\langle SMLS;N\, s|N\, n\right\rangle S_{1}
(k^{s}_{1})S_{2}(k^{s}_{2}),
\end{equation}
 with \( n \) and \( m \) even and the sum in \( s \) restricted to 
the positive
zeros of \( {\mathcal{H}}_{N}(x) \). The coefficients \( \left\langle SMLS;N\, 
s|N\, n\right\rangle  \)
are equal to \( \left\langle MLS;N\, s|N\, n\right\rangle  \) up to a factor
\( \sqrt{2} \) (see appendix). 

We remark the simplicity of the expression (\ref{eq:derived_Snm}) as compared
to the exact solution of the problem. Part of the payoff for this simplicity
is the existence of two undetermined parameters, \( \bar{v} \) and \( R_{0} \).
\emph{}The optimal value of \( R_{0} \) is nevertheless constrained by two
considerations. On one side, it can not be too large, as at large distances
the interaction between the two fragments would dominate over the interaction
due to the target, and so ignoring the correlations between the particles for
values just below \( R_{0} \) would not be a good approximation. On other side,
the value of \( R_{0} \) should be large enough to allow us to ignore the 
interaction
with the target for \( R \) above \( R_{0} \). Thus, the matching radius must
be about the size of the system. Moreover, \( \bar{v} \) must be of the order
of the expectation value of \( v(r) \) on the ground state.

\section{Comparison of USA and exact calculations \label{sec:apply}}
In this section we analyse the reliability of USA by comparing its predictions
with the exact results.
In Fig.\ \ref{Fig:Smat_usa} the S-matrix coefficients for the USA model 
(dot-dashed
line) with the phases \( \delta _{ns} \) given by (\ref{eq:delta_ns}) are
compared with the exact calculation (solid line) and the adiabatic approach
(dashed line). The calculations have been performed using the matching radius
\( R_{0}=0.6a_{0} \) and an average potential \( \bar{v}=0 \). These values
were determined by fitting the elastic S-matrix at high energies, where the
model is expected to be more accurate. We notice that for these energy 
independent
parameters a good description of the elastic and inelastic coefficients 
is achieved
for energies above \( \sim 6\, \hbar \omega  \). By contrast, the adiabatic
model does not seem to give a good description of the S-matrix at any 
scattering
energy.


A quantity closely related to the S-matrix coefficients is the \emph{average
final excitation energy}. It gives an idea on the degree of excitation of the
final system. It has been defined as

\begin{equation}
\left\langle \epsilon \right\rangle =\sum _{n=0}(\epsilon _{n}-\epsilon _{0})
|S_{0n}|^{2},
\end{equation}
 where the sum must be extended only to the set of open channels. 

\begin{figure}
{\par\centering \resizebox*{0.55\textwidth}{!}{\rotatebox{-90}
{\includegraphics{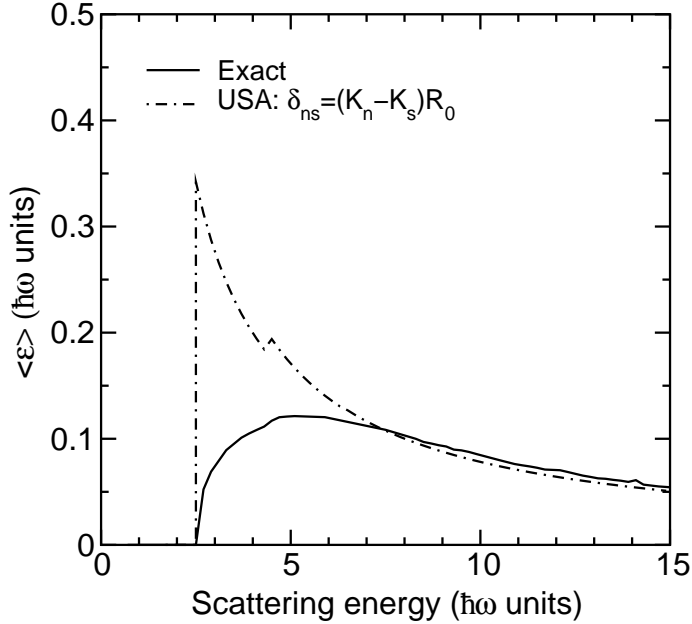}}} \par}
\caption{\label{Fig:<e>}Average excitation energy. The solid line 
corresponds to the
exact calculation. The dot-dashed line corresponds to the USA with 
the prescription
\protect\( \delta _{ns}=(K_{n}-\tilde{K}_{s})R_{0}\protect \).}
\end{figure}

In Fig.\ \ref{Fig:<e>} we show the energy dependence of the average excitation
energy, \( \langle \epsilon \rangle  \). As in the previous case, the USA model
(dot-dashed line) agrees well with the exact calculation (solid line) for 
energies
above \( 6\, \hbar \omega  \). On the contrary, the USA model, as expected,
does not describe properly the low energy regime. In particular, a spurious
discontinuity is observed for the threshold at 
\( E={5\over 2}\, \hbar \omega  \).

%


%

\begin{figure}
{\par\centering \resizebox*{0.65\textwidth}{!}{\rotatebox{-90}
{\includegraphics{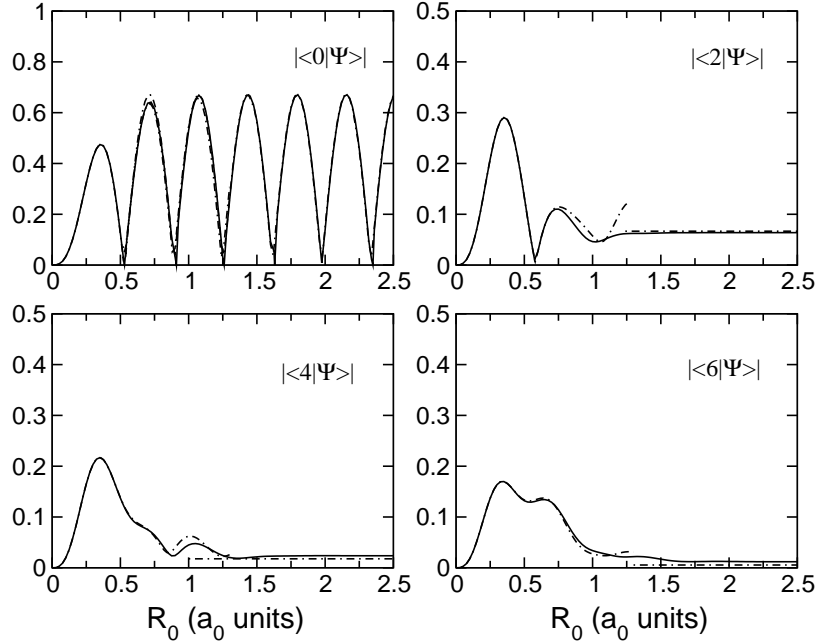}}} \par}
\caption{\label{Fig:<n|WF>_E10}Projection of the total wave function on 
the internal
eigenstates for a scattering energy \protect\( E=10\, \hbar \omega \protect \).
The solid line is the exact calculation, the dot-dashed line corresponds to
the formula (\ref{eq:derived_Snm}), with the prescription 
\protect\( \delta _{ns}=(K_{n}-\tilde{K}_{s})R_{0}\protect \). }
\end{figure}

In Fig.\ \ref{Fig:<n|WF>_E10} we compare the wave function given by the USA
approach with the exact calculation at \( E=10\, \hbar \omega  \). The curve
for the exact calculation has been obtained using expression 
(\ref{eq:<n|Phi>}).
In the USA model, represented by the dot-dashed line, one has to distinguish
the asymptotic and the interaction regions. For the asymptotic wave function,
the interaction with the wall is neglected (\( v_{i}(r_{i})=0 \)) and so the
range of values of \( r \) is unrestricted. Then, the projection on a state
\( n \) is simply given by 

\begin{equation}
\Psi ^{asym}_{n}(R)\rightarrow \delta _{n0}\frac{1}{\sqrt{v_{0}}}e^{-iK_{0}R}
-\frac{S_{0n}}{\sqrt{v_{n}}}e^{iK_{n}R}.
\end{equation}

By contrast, in the interacting region, explicit account is taken for the wall
and an expression similar to (\ref{eq:<n|Phi>}) should be used instead. Thus,
the total wave function in II (\ref{eq:WFii(r,R)}) is projected on the 
different
eigenstates, taking into account that the integration is restricted to the 
interval \( |r|<2R \). 


At this collision energy, Fig.\ \ref{Fig:Smat_usa} indicates that the 
scattering
coefficients are well reproduced by (\ref{eq:derived_Snm}) with the phases
\( \delta _{ns}=(K_{n}-\tilde{K}_{s})R_{0} \). Then, we have adopted this 
prescription
(with the same matching radius) to describe the approximated wave function at
this energy. As can be seen in Fig.\ \ref{Fig:<n|WF>_E10}, the agreement with
the exact calculation is quite good for all the states. 

As a general rule, our calculations show a better agreement for the elastic
and first excited states, and it tends to be worse for excited states 
of increasing
energies. This is expected because within the USA model we restrict the basis
to the set of open channels and so, the effect of this truncation
becomes more evident as we explore excited states close to the cut off.

\begin{figure}
{\par\centering \resizebox*{0.65\textwidth}{!}{\includegraphics
{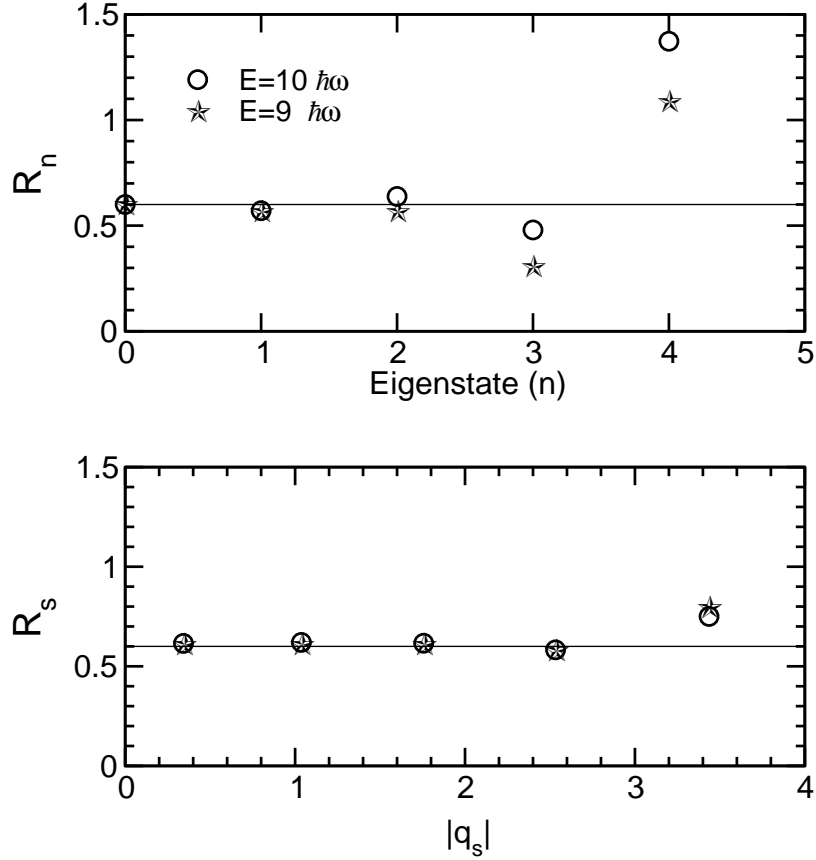}} \par}
\caption{\label{Fig:RnRs_e9-10}Behaviour of the ratios 
\protect\( R_{n}=\gamma _{n}/K_{n}\protect \)
(upper figure) and \protect\( R_{s}=\chi _{s}/\tilde{K}_{s}\protect \) (lower
figure) versus the value of \protect\( n\protect \) and 
\protect\( |q_{s}|\protect \),
respectively, for the scattering energies 
\protect\( E=9\, \hbar \omega \protect \)
(stars) and \protect\( E=10\, \hbar \omega \protect \) (circles). The phases
\protect \( \gamma _{n}\protect \) and \protect \( \chi _{s}\protect \) have
been calculated by fitting the S-matrix elements 
predicted by (~\ref{eq:derived_Snm})
to the exact calculation.}
\end{figure}

We have explored in more detail the limits of the USA model as well as 
the validity
of the prescription (\ref{eq:delta_ns}) for the phases. In order to do that
we start from the exact expression of the S-matrix for a fixed 
collision energy.
Making use of (\ref{eq:derived_Snm}), we fit the phases \( \gamma _{n} \)
and \( \chi _{s} \) appearing in \( \delta _{ns} \) in order to reproduce
the exact calculation. We find that the S-matrices can be accurately fitted
for all the scattering energies (even for very small values). Although these
phases, \( \gamma _{n} \) and \( \chi _{s} \), can not be written exactly
as \( K_{n}R_{0} \) and \( \tilde{K}_{s}R_{0}, \) respectively, one can 
define effective
energy and channel dependent radii so that \( R_{n}=\gamma _{n}/K_{n} \) and
\( R_{s}=\chi _{s}/\tilde{K}_{s} \). In Fig.\ \ref{Fig:RnRs_e9-10} we plot
the values of the quantities \( R_{n} \) versus \( n \)
and \( R_{s}=\chi _{s} \) versus the absolute value of \( q_{s} \),
i.e., the internal momentum at which the state \( \langle q|MLS;Ns\rangle  \)
is peaked. The selected collision energies are \( E=9\, \hbar \omega  \) and
\( E=10\, \hbar \omega  \). In both cases we deal with a total of 
five HO states,
namely, the ground state and the first four excited states with even parity.
Using the prescription (\ref{eq:delta_ns}) both quantities are just \( R_{0} \)
for all values of \( n \) or \( s \). Note that the value of \( R_{s} \)
is rather constant and very close to the matching radius used in our 
calculations,
i.e., \( R_{0}=0.6a_{0} \). This constant value has been also plotted in the
figure for reference. The values of \( R_{n} \) are also very close to 
\( R=R_{0} \)
for the lower values of \( n \), but they tend to deviate from our prescription
for values of \( n \) close to the threshold. It is remarkable that the values
of \( R_{n} \) and \( R_{s} \) are mostly independent on the scattering energy
and of the individual state considered. This indicates that for scattering 
energies
large compared to \( \hbar \omega  \), the USA works very well, and the 
matching
radius can be taken as a constant, related to the size of the system, 
and independent
on the energy or the internal state. For lower energies, the USA may still be
used, but in this regime the radius \( R_{0} \) depends on the energy and the
state considered.

\section{Conclusions\label{sec:conclusions}}

We have formulated a one dimensional problem consisting on the scattering of
two particles, interacting with a HO potential, that collide with an infinite
potential. This problem contains some of the dynamical features of the 
scattering of composite systems in atomic, molecular and nuclear physics, 
which interact with a  target through short range interactions.

We have obtained the \emph{exact} solution of the problem using two different
procedures. The first one consists of imposing the adequate boundary conditions
on the scattering wave function. The second procedure deals with a basis of
configuration localized states (CLS), which are wave functions with a strong
spatial localisation. Both procedures converge, provided that a sufficiently
large basis of states is used. The main characteristic of the exact solution
is that, for large scattering energy, the elastic scattering dominates. 
In order
to achieve convergence, the inclusion of closed channels, i.e., states with
internal energy larger than the scattering energy, is required. Although these
states do not contribute to the S-matrix they must be taken into account in
the calculations to obtain accurate results. This fact indicates that 
for short-range
interactions one should be very careful when truncating the basis of states
used in continuum discretized calculations. 

We have compared our \emph{exact} results with the adiabatic approach, that
considers the relative coordinate frozen during the scattering process. The
results disagree completely. This indicates that the adiabatic approximation
could be inaccurate when the interactions of the fragments with the target
have a very short range.
These short range interactions could couple to highly excited internal states 
for which the adiabatic approximation is not valid. 
However, it should be reminded that our calculations make use of a sharp 
infinite wall. Thus, they will be relevant when the range of the interaction
is short  compared not only with the size of the projectile but also 
with the wavelength describing the motion of the fragments with respect to 
the target.

We have developed a model which describes the scattering of a composite object
in terms of the scattering wave functions and the S-matrices of the fragments.
The model, that we call Uncorrelated Scattering Approximation (USA), neglects
the correlations between the fragments during the scattering. The
application of the USA to our model problem gives
an expression for the S-matrix of the composite system in terms of the product
of the S-matrices of the fragments. Similarly, the scattering wave function
is given as a combination of the product of regular wave functions of the fragments.
The particular superposition is determined by application of the asymptotic
boundary conditions. The application of the USA to our model problem 
relies on the use of two parameters. The most important one is the distance 
\( R_{0} \) at which the asymptotic and uncorrelated wave functions are 
matched. We have fixed this value to $0.6 a_0$, in terms of the oscillator 
length, for all our calculations. It should be noticed that this parameter 
has a similar meaning to the matching radius that is used in R-matrix theory.
The other parameter is the average 
potential \( \bar{v} \) that replaces the interaction between the fragments. 
We have set this value to $\bar v = 0$.

By considering the matching radius to be energy- and state-dependent, the exact
S-matrix and scattering wave functions are accurately reproduced. Moreover,
for high scattering energies the elastic and inelastic S-matrices are well 
reproduced
by taking a fixed value of the matching radius. 

In general, the agreement with the exact calculation is better for the 
observables
associated with the ground and first excited states and they tend to be worse
for states with excitation energies close to the total energy. This is 
attributed
to the fact that for these states the relative velocity between the fragments
is small and then the correlations are expected to be more important.

The one-dimensional model presented in this work can be a useful test case
to check the validity of different approaches used in the description of
the scattering of
composite systems. The present choice of a sharp  infinite wall for
the description of the interaction with the target, and harmonic oscillator
for the interactions between the fragments has the advantage that sets
$\hbar \omega$ as the unique scale for energies and $a_0$ as the unique scale
for lengths. In this sense, our results, which are expressed in those units, 
are valid for any value of the mass or harmonic constant.
However, the model could be done more realistic, and more
complicated, by substituting the 
infinite wall for an exponential function, and substituting the harmonic 
oscillator by finite potentials. 

\section*{Acknowledgements }

We are grateful to Prof. R. C. Johnson for his worthy comments on the adiabatic
model. A.M.M. acknowledges a research grant from the Fundaci\'on C\'amara of the
University of Sevilla. This work has been partially supported by the Spanish
DGICyT, project PB98-1111.



%


\newpage
\section* {Appendix A}

In this work different sets of basis states have been used, based on harmonic
oscillator (HO) wave functions. These obey the general form

\begin{equation}
\label{eq:wfosc}
\phi _{n}(r)={\mathcal{N}}_{n}^{-\frac{1}{2}}\exp \left( -\frac{r^{2}}
{2a^{2}_{0}}\right) {\mathcal{H}}_{n}
\left( \frac{r}{a_{0}}\right) \, \, ;\, \, 
\, n=0,1,\ldots 
\end{equation}
 where \( a_{0}=\sqrt{\hbar /\mu \omega } \) is the oscillator length, \( 
{\mathcal{N}}_{n} \)
a normalisation constant and \( {\mathcal{H}}_{n} \) the Hermite polynomial of
order \( n \). In all the calculations we restrict the infinite set of states
to a finite family of \( N \) states, denoted by \( \{|N\, n\rangle ;\, n=0,
\ldots ,N-1\} \),
where \( \langle r|N\, n\rangle =\phi _{n}(r) \). 

By diagonalizing the position operator in the truncated basis of HO wave 
functions,
a new set of \( N \) states is obtained which have the property of being 
localised
in configuration space. They are called Configuration Localised States (CLS)
and are denoted by \( \{|CLS;Ns\rangle ;s=1,\ldots ,N\} \). These two families
of states are related by means of the orthogonal transformation

\begin{equation}
\label{eq:CLSbis}
|CLS;Ns\rangle =\sum ^{N-1}_{n=0}\left\langle Nn|CLS;Ns\right\rangle 
|Nn\rangle 
\end{equation}

For HO wave functions, the state \( \langle r|CLS;Ns\rangle  \) is localised
around \( r=a_{0}x_{s} \), where \( x_{s} \) is the \( s \)-th zero of the
Hermite polynomial \( {\cal {H}}_{N}(x) \). In this case, the transformation
coefficients are given by (see~\cite{Per99})

\begin{equation}
\left\langle Nn|CLS;Ns\right\rangle =\left[ \frac{2^{N-n}}{2N}\frac{(N-1)!}
{n!}\right] ^{1/2}\frac{\mathcal{H}_{n}(x_{s})}{\mathcal{H}_{N-1}(x_{s})}.
\end{equation}

In analogy with the CLS, it is possible to define internal states localised
in momentum space, known as Momentum Localized States (MTS). This is carried
out by diagonalizing the momentum operator in the truncated HO basis. Thus,
starting with the basis of \( N \) states, this procedure provides a new set
of \( N \) internal states, each one of them is peaked around a certain 
momentum.
As in the case of the CLS there is an orthogonal transformation relating both
sets of states:

\begin{equation}
\label{eq:MLS}
|MLS;Ns\rangle =\sum ^{N-1}_{n=0}\langle Nn|MLS;Ns\rangle |Nn\rangle ,
\end{equation}
 where \( |MLS;Ns\rangle  \) represents a MLS. Then \( \langle q|MLS;Ns\rangle
  \)
is localised around \( q=x_{s}/a_{0} \) where, due to the formal analogy of
the HO wave functions in momentum and configuration space, \( \{x_{s}\} \)
are again the zeros of the Hermite polynomial \( {\mathcal{H}}_{N} \). This 
analogy
provides also a simple relation between the transformation coefficients 
\( \langle MLS;Ns|Nn\rangle  \)
and \( \langle CLS;Ns|Nn\rangle  \). In this work, we take the internal wave
function to be real in configuration space, and so they will be affected by
a factor \( (-i)^{n} \) in momentum space. Moreover, we take the CLS to be
real functions, so the coefficients \( \langle CLS;Ns|Nn\rangle  \) are real
numbers. Therefore, 

\begin{equation}
\langle MLS;Ns|Nn\rangle =(-1)^{n}\langle CLS;Ns|Nn\rangle .
\end{equation}

Although the formalisms presented in this work do not require the fragments
to have equal masses, we have performed all the calculations for this 
particular
situation. In this case, parity conservation implies that only positive parity
states are suitable to be populated during the process. In fact, the 
coefficients
\( S_{0n} \) derived from the exact calculations of sections 
(\ref{sec:exact_direct})
and (\ref{sec:exact_CLS}) are found to be zero for odd values of \( n \).
This is also satisfied by eq.\ (\ref{eq:derived_Snm}), as can be easily 
verified. 

Under these considerations one is allowed to exclude those inhibited states
from the beginning. This permits to work with the truncated basis of states:
\{\( \phi _{n}(r);\, n=0,2,\ldots ,N-2\} \), with \( N \) even. This requires,
however, some care in the evaluation of the localised states in the new basis.
The formalism of CLS can not be directly applied to this set, as many of its
properties entails the sum over both even and odd states. The starting point
for the construction of the CLS formalism requires a set of functions of the
form~\cite{Per99}

\begin{equation}
\label{eq:psi}
\psi _{m}(x)=\langle x|j\, m\rangle ={\mathcal{N}}^{-1/2}_{jm}F(y)
{\mathcal{P}}_{m}(y),\, \, \, \, \, \, \, m=0,1,\ldots ,j-1,
\end{equation}
 where \( y \) is a certain function of \( r \), \( {\mathcal{N}}_{jm} \) a
normalisation constant, \( F \) an arbitrary function of \( y \) and \( 
{\mathcal{P}}_{m} \)
is a polynomial of order \( m \). 

The drawback outlined above can be overcome in the case of HO wave functions
by writing the Hermite polynomials in terms of generalised Laguerre functions
(see, for instance, Ref.~\cite{Abr72}):

\begin{equation}
\label{eq:phi2n}
\phi _{2m}(r)=(-1)^{m}m!2^{2m}{\mathcal{N}}^{-\frac{1}{2}}_{2m}\exp 
\left( -\frac{r^{2}}{2a^{2}_{0}}\right) {\mathcal{L}}^{(-1/2)}_{m}(y),\, \, \, 
y=\left( \frac{r}{a_{0}}\right) ^{2}.
\end{equation}

Then, taking \( {\mathcal{P}}_{m}(y)\equiv {\mathcal{L}}^{(-1/2)}_{m}(y) \), 
\( F(y)\equiv \exp (-y/2) \)
and \( j\equiv \frac{N}{2} \) we can identify \( \psi _{m}(x)\equiv 
\phi _{2m}(r) \)
(\( m=0,1,\ldots ,\frac{N}{2}-1 \)). The set of configuration localised states
are now calculated for the new set of functions. It requires the calculation
of the roots of the polynomial \( {\mathcal{L}}^{(-1/2)}_{N/2}(y) \) which, 
attending
to (\ref{eq:phi2n}), are just the square of the zeros of 
\( {\mathcal{H}}_{N}(x) \).
The new set of localised states are given by a linear combination of the states
\( \phi _{n}(r) \) and so they are even functions with respect to the variable
\( r \). Therefore, we call them \emph{}Symmetric Configuration Localised States,
SCLS. The transformation between the set of states (\ref{eq:phi2n}) and the
SCLS is then expressed as

\begin{equation}
\label{eq:CLS_Lag}
|SCLS;Ns\rangle =\sum ^{N-2}_{n=0(even)}\left\langle N\, n|SCLS;\, N\, s
\right\rangle |N\, n\rangle ,\, \, s=1,...,\frac{N}{2}.
\end{equation}

The state \( \langle r|SCLS;Ns\rangle  \) is localised around \( r=\pm a_{0}
x_{s} \)
where \( \{x_{s}\} \) are the positive zeros of \( {\mathcal{H}}_{N}(x) \).

The coefficients \( \left\langle SCLS;\, N\, s|N\, n\right\rangle  \) are found
to be equal to those appearing in eq.\ (\ref{eq:CLSbis}) up to a factor 
\( \sqrt{2} \): 

\begin{equation}
\left\langle SCLS;N\, s|N\, n\right\rangle =\sqrt{2}\left\langle CLS;N\, 
s|N\, n\right\rangle 
\end{equation}
 for \( n \) even. 

\begin{figure}
{\par\centering \resizebox*{0.45\textwidth}{!}
{\includegraphics{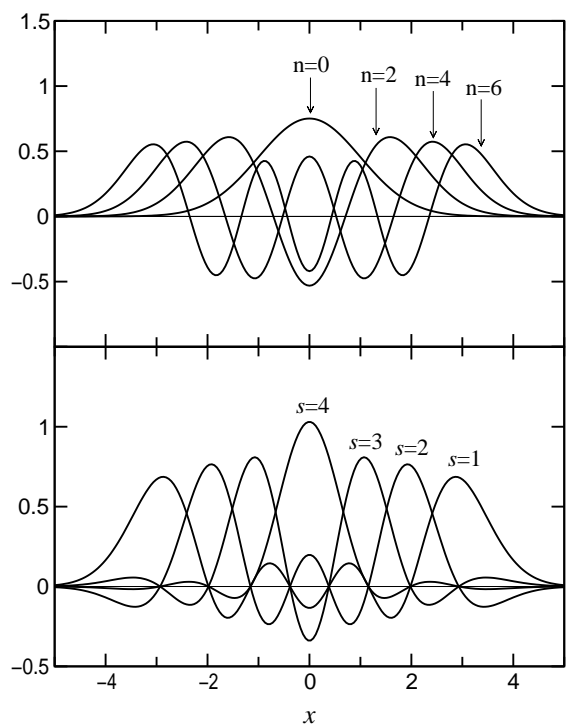}} \par}
\caption{\label{Fig:CLS}Harmonic oscillator wave functions corresponding 
to \protect\( n=0,2,4\protect \)
and 6 (upper figure) and associated Symmetric Localized States, labeled with
the index \protect\( s\protect \). }
\end{figure}

As an example, in Fig.\ \ref{Fig:CLS} the set of HO wave functions 
\( n=0,2,4,6 \)
are plotted versus the adimensional variable \( x \) (upper figure). 
The corresponding
Symmetric Localised States are also plotted (lower figure) and labelled with
the index \( s \) (\( s=1,2,3,4 \)). Notice that each one of these localised
states is peaked around two symmetrical points, corresponding to symmetrical
roots of the Hermite polynomial \( {\mathcal{H}}_{8}(x) \). 

In a similar way, it is possible to construct Symmetric Momentum Localised 
States\emph{,}
which are given by means of the transformation 

\begin{equation}
\label{eq:SMLS}
|SMLS;Ns\rangle =\sum ^{N-2}_{n=0(even)}\left\langle N\, n|SMLS;\, N\, 
s\right\rangle |N\, n\rangle ,\, \, s=1,...,\frac{N}{2}.
\end{equation}

The state \( \langle q|SCLS;Ns\rangle  \) is localised around 
\( q=\pm x_{s}/a_{0} \)
where \( \{x_{s}\} \) are again the positive zeros of 
\( {\mathcal{H}}_{N}(x) \).
In the case of the HO basis, the transformation coefficients are related to
those in configuration space:

\begin{equation}
\left\langle SMLS;N\, s|N\, n\right\rangle =\sqrt{2}(-i)^{n/2}\left\langle 
CLS;N\, s|N\, n\right\rangle .
\end{equation}

\end{document}